\documentclass[a4paper,12pt]{article}

\usepackage{amsmath}
\usepackage{amsfonts}
\usepackage{graphicx} 

\numberwithin{equation}{section}

\newcommand{\al}{{\alpha}}
\newcommand{\del}{{\delta}}
\newcommand{\Del}{{\Delta}}
\newcommand{\eps}{{\varepsilon}}
\newcommand{\kap}{{\kappa}}
\newcommand{\la}{{\lambda}}
\newcommand{\La}{{\Lambda}}
\newcommand{\om}{{\omega}}
\newcommand{\sig}{{\sigma}}
                
\newcommand{\Ocal}[1]{{\mathcal{O}^{(#1)}}}
\newcommand{\Hcal}{{\mathcal{H}}}
\newcommand{\Bcal}{{\mathcal{B}}}
\newcommand{\Jcal}{{\mathcal{J}}}
\newcommand{\Acal}{{\mathcal{A}}}
\newcommand{\Rcal}{{\mathcal{R}}}
\newcommand{\Lcal}{{\mathcal{L}}}
\newcommand{\Dcal}{{\mathcal{D}}}
\newcommand{\Dcalh}{{\mathcal{\hat{D}}}}

\newcommand{\nn}{{\nonumber}}
\newcommand{\pd}{{\partial}}
\newcommand{\pard}[1]{{\frac{\pd}{\pd #1}}}
\newcommand{\eqand}{{\qquad \mathrm{and} \qquad}}

\newcommand{\etal}{{\emph{et al.}\ }}
\newcommand{\Rbb}{{\mathbb{R}}}
\newcommand{\lak}[1]{{\la_{\kap ,m}^\mathrm{#1}}}
\newcommand{\lakz}[1]{{\la_{\kap ,0}^\mathrm{#1}}}

\newcommand{\lb}{{\ell}}
\newcommand{\nb}{{n}}

\newcommand{\Om}{\Omega} 
\newcommand{\vphi}{\varphi} 
\newcommand{\Phis}{\Phi^\mathrm{S}} 
\newcommand{\Phia}{\Phi^\mathrm{A}} 

\DeclareTextFontCommand{\textwasy}{\wasyfamily}
\def \wasyfamily{\fontencoding{U}\fontfamily{wasy}\selectfont}
\def \thorn{{\wasyfamily\char105}}
\DeclareTextCommand{\dh}{OT1}{{\wasyfamily\char107}}
\newcommand{\tho}{{\textrm\thorn}}
\newcommand{\eth}{{\textrm{\dh}}}

\newcommand{\DelLA}{{\Delta_\mathrm{L}^\mathcal{A}}}

\newcommand{\nablah}{{\hat{\nabla}}}
\newcommand{\gh}{{\hat{g}}}

\newcommand{\Rh}{{\hat{R}}}
\newcommand{\Ih}{{\hat{I}}}
\newcommand{\Jh}{{\hat{J}}}

\newcommand{\Ah}{{\hat{A}}}
\newcommand{\Bh}{{\hat{B}}}

\newcommand{\alh}{{\hat{\alpha}}}
\newcommand{\betah}{{\hat{\beta}}}
\newcommand{\gammah}{{\hat{\gamma}}}
\newcommand{\delh}{{\hat{\delta}}}

\newcommand{\Ybb}{{\mathbb{Y}}}
\newcommand{\Id}{{\mathbf{1}}}
\newcommand{\Acalh}{{\mathcal{A}}}
\newcommand{\CP}[1]{{\mathbb{CP}^{#1}}}
\newcommand{\Jcalh}{{\mathcal{J}}}
\newcommand{\Pcalh}{{\mathcal{P}}}

\newcommand{\be}{\begin{equation}}
\newcommand{\ee}{\end{equation}}

\title{Perturbations of near-horizon geometries and \\ instabilities of Myers-Perry black holes}
\author{Mark N.~Durkee and Harvey S.~Reall\\
        {\small\it DAMTP, University of Cambridge, Centre for Mathematical Sciences,}\\
        {\small\it Wilberforce Road, Cambridge, CB3 0WA, United Kingdom}\\
        {\small\tt M.N.Durkee@damtp.cam.ac.uk, H.S.Reall@damtp.cam.ac.uk}}

\begin{document}
\maketitle
\begin{abstract}
It is shown that the equations governing linearized gravitational (or electromagnetic) perturbations of the near-horizon geometry of any known extreme vacuum black hole (allowing for a cosmological constant) can be Kaluza-Klein reduced to give the equation of motion of a charged scalar field in $AdS_2$ with an electric field. One can define an effective Breitenl\"ohner-Freedman bound for such a field. We conjecture that if a perturbation preserves certain symmetries then a violation of this bound should imply an instability of the full black hole solution.  Evidence in favour of this conjecture is provided by the extreme Kerr solution and extreme cohomogeneity-1 Myers-Perry solution. In the latter case, we predict an instability in seven or more dimensions and, in 5d, we present results for operator conformal weights assuming the existence of a CFT dual. We sketch a proof of our conjecture for scalar field perturbations.

\end{abstract}

\section{Introduction}\label{sec:intro}
The study of linearized gravitational perturbations of the Kerr spacetime is tractable because the equations governing such perturbations can be decoupled and reduced to a wave equation for a single complex scalar \cite{Teukolsky}.  No instabilities are found \cite{Press:1973zz,Stewart:1975vg,Whiting:1988vc}.  Following a prediction in Ref.\ \cite{Emparan:2003sy}, it has been demonstrated recently that certain higher-dimensional rotating black holes suffer linearized gravitational instabilities \cite{Dias:2009iu,Dias:2010eu,Dias:2010maa,Dias:2010gk}.\footnote{
A different kind of instability has been found numerically \cite{shibata}.}. To extend these results to more general black holes, it would be very useful to have a decoupled equation describing gravitational perturbations. By this, we mean a linear partial differential equation in which the unknown is a gauge-invariant quantity with the same number of degrees of freedom as the gravitational field. 

This problem was investigated in Ref.\ \cite{decoupling}.  It was shown that, for perturbations of an algebraically special spacetime in $d$ dimensions (e.g.\ a Myers-Perry \cite{mp} black hole), one can introduce a $(d-2) \times (d-2)$ traceless symmetric matrix $\Omega_{ij}$ that is linear in the metric perturbation and invariant under both infinitesimal coordinate transformations and infinitesimal basis transformations. This object has the same number of degrees of freedom as the gravitational field. It is the higher-dimensional generalization of the Newman-Penrose scalar $\Psi_0$ that satisfies a decoupled equation of motion in the Kerr geometry (or any other algebraically special vacuum solution). Therefore it is a natural object to consider in any study of gravitational perturbations. 

In detail, $\Omega_{ij}$ is defined as follows. In the perturbed spacetime, introduce a null basis $\{\ell,n,m_i\}$ where $\ell$ and $n$ are null, and $m_i$ are orthonormal spacelike vectors orthogonal to $\ell$ and $n$. Then $\Omega_{ij} =C_{abcd} \ell^a m_i^b \ell^c m_j^d$ where $C_{abcd}$ is the Weyl tensor. The gauge-invariance properties just mentioned hold if, in the unperturbed spacetime, $\ell$ reduces to the `multiple Weyl aligned null direction' associated to the algebraically special property.\footnote{The concept of a Weyl aligned null direction (WAND) was defined in Ref.\ \cite{cmpp}.  An understanding of this classification is not required for reading this paper, except for the calculations in Appendix \ref{app:nhframe}.}

Unfortunately, Ref.\ \cite{decoupling} found that $\Omega_{ij}$ does not satisfy a decoupled equation of motion in most algebraically special spacetimes. It was shown that the necessary and sufficient conditions for decoupling are that the spacetime should admit a null geodesic congruence with vanishing expansion, rotation and shear. A spacetime admitting such a congruence is called a Kundt spacetime. Black hole solutions are not Kundt spacetimes. However, it was observed in Ref.\ \cite{decoupling} that the \emph{near-horizon (NH) geometry} of any extreme black hole (BH) solution {\it is} a Kundt spacetime.  Hence the decoupled equation of Ref.\ \cite{decoupling} provides a natural starting point for a study of linearized gravitational perturbations of near-horizon geometries.

The purpose of the present paper is to initiate a study of gravitational perturbations of near-horizon geometries of extreme black holes.  The results of Ref.\ \cite{decoupling} apply to all vacuum spacetimes (allowing for a cosmological constant) and so the extreme black holes in question include Myers-Perry black holes \cite{mp} and doubly-spinning black rings \cite{Pomeransky}.  The reason for our interest in perturbations of near-horizon geometries is the following question: can one learn something about stability of an extreme black hole solution from a study of perturbations of its near-horizon geometry? Clearly, we will not be able to deduce that the full black hole is stable just by looking at its near-horizon geometry. So, a more precise question is: \emph{does an instability of the near-horizon geometry imply an instability of the full black hole solution?}

If the answer is yes then this would give a fairly simple way of predicting instabilities of extreme black holes because perturbations of a near-horizon geometry can be studied using the decoupled equation of Ref.\ \cite{decoupling}. Furthermore, if an extreme black hole is unstable then it is likely that near-extreme black holes also are unstable.

We should clarify what we mean by an instability of a near-horizon geometry. As we shall review below, any known near-horizon geometry takes the form of a compact space $\Hcal$ fibred over $AdS_2$. It can be regarded as a Kaluza-Klein (KK) compactification with internal space $\Hcal$. In section  \ref{sec:decoupling} we shall show that the decoupled equation governing linearized gravitational (or scalar field or electromagnetic) perturbations can be `KK reduced' by expanding in eigenfunctions of a certain operator on $\Hcal$. This is non-trivial because the `KK gauge fields' arising from rotation of the black hole are non-vanishing.  The result is a massive, charged, scalar field in $AdS_2$ with a homogeneous electric field. One can define an `effective Breitenl\"ohner-Freedman (BF) bound' \cite{BF} for such a field.  We shall say that the near-horizon geometry is unstable if there is some mode that violates this bound.

Some motivation for believing that an instability of a near-horizon geometry implies an instability of the full black hole comes from studies of charged scalar fields in the background of an extreme Reissner-Nordstrom-AdS black hole.  Numerical results \cite{Hartnoll:2008kx,Denef:2009tp} suggest that the scalar field becomes unstable in the black hole geometry when the near-horizon $AdS_2$ BF bound is violated.  In fact instability can occur even for an uncharged scalar field.  In this case, we shall present a proof (in Section \ref{sec:nhinstab}) that instability of the near-horizon geometry does imply instability of the full black hole.

Returning to gravitational perturbations, what happens for $d=4$? The near-horizon extreme Kerr (NHEK) geometry was introduced in Ref.\ \cite{Zaslavsky:1997uu,Bardeen:1999px}. It has $\Hcal=S^2$ (with an inhomogenous metric).  Linearized gravitational perturbations of NHEK were studied in Refs.\ \cite{Amsel:2009ev,Dias:2009ex}. After KK reduction to $AdS_2$, it turns out that certain non-axisymmetric modes violate the effective BF bound. In this sense, the NHEK geometry is unstable against linearized gravitational perturbations. But the full Kerr solution is believed to be stable. 

What does this example teach us? We could conclude that instability of the near-horizon geometry does not imply instability of the full black hole. Instead, we shall argue that instability of the near-horizon geometry {\it does} imply instability of the full black hole {\it if the unstable mode respects certain symmetries}. In the Kerr example, the symmetry in question is axisymmetry. Axisymmetric perturbations of NHEK {\it do} respect the BF bound; the stability of such modes is consistent with the stability of the full black hole. 

Before attempting to understand why an instability of the near-horizon geometry implies an instability of the full black hole when certain symmetries are respected, we will start by gathering some more data. In section \ref{sec:nhmp}, we will consider the most symmetric rotating black hole solutions: Myers-Perry (MP) black holes \cite{mp} in an odd number of dimensions, with equal angular momenta.  Such black holes are cohomogeneity-1 (i.e.\ the metric depends non-trivially on just one coordinate). The Killing field tangent to the horizon generators has the form ${\bf k} + \Omega_H {\bf m}$ where ${\bf k}$ is the generator of asymptotic time translations, ${\bf m}$ is an angular Killing field with closed orbits, and $\Omega_H$ is the angular velocity of the black hole.

In the extreme limit, such a black hole has a homogeneous near-horizon geometry for which $\Hcal=S^{d-2}$ (with a homogeneous metric). After KK reduction to $AdS_2$, we find that there exist modes that violate the effective BF bound, but most of these violate the symmetry generated by ${\bf m}$. These are the analogue of the non-axisymmetric modes in NHEK. What about modes that preserve the symmetry generated by ${\bf m}$? For $d=5$, we find that such modes always respect the BF bound, just as for NHEK. However, for $d \ge 7$, we find that some of these modes violate the BF bound.  

How does this compare with the stability properties of the full extreme black hole solution? Refs.\ \cite{Kunduri:2006,Murata:2008,Dias:2010eu} studied gravitational perturbations of {\it non-extreme} cohomogeneity-1 MP solutions. It is natural to expect that a reliable guide to the stability of an extreme black hole should be the stability of black holes that are very close to extremality. For modes that are invariant under the symmetry generated by ${\bf m}$, it turns out that, in the cases for which data exists, for any mode that is unstable in the near-horizon geometry, there is a corresponding unstable mode of the full black hole solution close to extremality. This leads us to predict confidently that {\it all} cohomogeneity-1 MP black holes with $d \ge 7$ are unstable sufficiently close to extremality. 

Are these isolated examples of a more general result?  If so, under what circumstances does an instability of the near-horizon geometry imply an instability of the full black hole?  The evidence from NHEK and cohomogeneity-1 black holes suggests that the instability should preserve a symmetry related to a rotational symmetry of the background geometry, so we will say a few words about such symmetries.

In any number of dimensions, the rigidity theorem guarantees the existence of at least one rotational Killing vector for any stationary black hole solution \cite{Hollands:2006rj}.  All {\it known} exact black hole solutions have more symmetry than this; they have multiple rotational symmetries. Consider a stationary black hole with $n$ commuting angular Killing fields $\partial/\partial \phi^I$ and a metric of the form
\begin{equation}
\label{ADM}
 ds^2 = - N(x)^2 dt^2 + g_{IJ}(x)\left(d\phi^I + N^I(x) dt\right)\left(d\phi^J + N^J(x) dt\right) 
        + g_{AB}(x) dx^A dx^B
\end{equation}
where $1 \le I,J \le n$, $\phi^I \sim \phi^I + 2\pi$, and the metric depends only on the coordinates $x^A$.  Known black hole solutions, e.g.\ Myers-Perry or black rings, have metrics of this form.

We can now present a conjecture for the circumstances under which we think a near-horizon geometry instability implies an instability of the full black hole:

\medskip \noindent 
{\bf Conjecture}.  {\it Consider linearized gravitational perturbations of the near-horizon geometry of an extreme vacuum black hole with metric \eqref{ADM}. Fourier decompose into modes with $\phi^I$ dependence $e^{im_I \phi^I}$. A sufficient condition for instability of the full black hole geometry is that the near-horizon geometry is unstable against perturbations with
}
  \begin{equation}\label{eqn:axicondition}
    m_I N^I(x) = 0
  \end{equation}
\medskip
We emphasize that $N^I(x)$ refers to the full black hole metric, not the near-horizon geometry.
For most MP black holes, or doubly-spinning black rings, the functions $N^I(x)$ are linearly independent so (\ref{eqn:axicondition}) implies $m_I = 0$ for all $I$. However, for MP solutions with enhanced symmetry this condition is less restrictive, e.g, in the cohomogeneity-1 case it implies only that $\Sigma_I m_I = 0$, which is equivalent to the perturbation being invariant under the symmetry generated by ${\bf m}$. 

How could one prove this conjecture? In section \ref{sec:nhinstab}, we sketch a proof that an instability of a {\it scalar field} in the near-horizon geometry of an extreme black hole implies an instability in the full black hole spacetime if the condition \eqref{eqn:axicondition} holds. This explains why the conjecture should be true for a scalar field. For gravitational perturbations, we do not have a complete argument but the results discussed above, and further evidence that we shall discuss, suggests that an argument similar to the scalar field case should also apply.

A final topic discussed briefly in this paper (section \ref{sec:kerrcft}) concerns the conjectured Kerr-CFT correspondence \cite{kerrcft}. It has been suggested that this extends to extreme Myers-Perry black holes \cite{popekerrcft}. For cohomogeneity-1, we predict an instability if $d \ge 7$ so $d=5$ seems the best-motivated case to consider. If we follow the usual AdS/CFT rules we can determine conformal weights of operators in the dual CFT using our results for gravitational perturbations of the near-horizon geometry. We find that all operators dual to gravitational perturbations respecting (\ref{eqn:axicondition}) have {\it integer} conformal weights (the same is true for NHEK). This is surprising (e.g. it would not be true for $d \ge 7$) and suggests that some symmetry is protecting the weights of these operators.

This paper is organized as follows. Section \ref{sec:decoupling} reviews near-horizon geometries and shows how the equations governing scalar field, gravitational, and electromagnetic perturbations can be Kaluza-Klein reduced to the equation of a scalar field in $AdS_2$. Section \ref{sec:nhmp} carries out this reduction for the case of cohomogeneity-1 Myers-Perry black holes and shows how the results are consistent with the above conjecture. Finally, section \ref{sec:nhinstab} gives further discussion of the conjecture. The Appendices contain details of our calculations. 

\section{Decoupling and near-horizon geometries}\label{sec:decoupling}
\subsection{Near-horizon geometries} \label{sec:nhgeom}

Consider an extreme black hole, i.e., one with a degenerate Killing horizon.  Gaussian null coordinates $(v,r,x^\mu)$ can be introduced (see e.g.\ \cite{Moncrief:1983,Friedrich:1998wq,Reall:2002bh}) in a neighbourhood of the horizon so that the metric takes the form
\begin{equation}
ds^2 = -r^2 F(r,x) dv^2 + 2 dv dr + 2 r h_\mu (r,x) dv dx^\mu 
       + \gamma_{\mu\nu} (r,x) dx^\mu dx^\nu .
\end{equation}
The Killing field tangent to the horizon is $\pd/\pd v$. The null vector field $n=\pd/\pd r$ is tangent to a congruence of null geodesics transverse to the horizon, which is at $r=0$. The functions $F$, $h_\mu$ and $\gamma_{\mu\nu}$ are smooth functions of $r$, with 
\begin{equation}
  F(r,x)=F(x) + {\cal O}(r), \qquad
  h_\mu(r,x)=h_\mu(x) + {\cal O}(r), \qquad 
  \gamma_{\mu\nu}(r,x)=\gamma_{\mu\nu}(x) + {\cal O}(r)
\end{equation}
near the horizon $r=0$.  The fact that $g_{vv} = {\cal O}(r^2)$ follows from degeneracy of the horizon (i.e.\ vanishing surface gravity). The near-horizon limit is defined by rescaling $v \mapsto v/\epsilon$, $r \mapsto \epsilon r$ and taking the limit $\epsilon \rightarrow 0$ \cite{Reall:2002bh}.  This gives the near-horizon geometry 
\begin{equation}
ds^2 = -r^2 F(x) dv^2 + 2 dv dr + 2 r h_\mu (x) dv dx^\mu 
       + \gamma_{\mu\nu} (x) dx^\mu dx^\nu .
\end{equation}
A calculation reveals that $\pd/\pd r$ is tangent to a null geodesic congruence with vanishing expansion, rotation, and shear. The existence of such a congruence is the defining property of  a Kundt spacetime. Hence the near-horizon geometry of any extreme black hole is a Kundt spacetime.

It turns out that the near-horizon geometries of all {\it known} extreme vacuum black hole solutions have more symmetry than is manifest in the above metric \cite{Bardeen:1999px,Kunduri:2007vf,Figueras:2008qh,Kunduri:2008rs,Chow:2008dp}. These near-horizon geometries can be written as a fibration over $AdS_2$:
\begin{equation}
\label{nhgeom}
 ds^2 = L(y)^2 \left( - R^2 dT^2 + \frac{dR^2}{R^2} \right) 
        + g_{IJ}(y) \left( d\phi^I - k^I R dT \right) \left( d\phi^J - k^J R dT \right) 
        + g_{AB}(y) dy^A dy^B
\end{equation}
where $\pd/\pd \phi^I$, $I=1, \ldots, n$ are the rotational Killing vector fields of the black hole and $k^I$ are constants.  The metric in the first set of round brackets is the metric of $AdS_2$ (written here in Poincar\'e coordinates).  The coordinates $\phi^I$ have period $2\pi$.  The metric depends non-trivially only on the $d-n-2$ coordinates $y^A$.

A calculation reveals that the vector fields $\lb$ and $\nb$ dual to $-dT \pm dR/R^2$ are tangent to affinely parameterized null geodesics with vanishing expansion, rotation and shear.  Ref.\ \cite{decoupling} proposed that a spacetime admitting two independent null geodesic congruences with vanishing expansion, rotation and shear be called a {\it doubly Kundt} spacetime.  Such a spacetime is of algebraic type D in the sense of Ref.\ \cite{cmpp}. If we consider perturbing such a spacetime then the perturbations in both $\Omega_{ij}$ and $\Omega'_{ij} \equiv C_{abcd} n^a m_i^b n^c m_j^d$ are gauge invariant and satisfy the decoupled equations given in Ref.\ \cite{decoupling}.

\subsection{Decomposition of perturbations}\label{sec:decomposition}

In this paper, we shall study scalar field, gravitational and electromagnetic perturbations of vacuum spacetimes of the form \eqref{nhgeom} (allowing for a cosmological constant: $R_{ab} = \Lambda g_{ab}$).  This includes, for example, the near-horizon geometries of extreme Myers-Perry-(AdS) black holes \cite{mp,Hawking:1998kw,Gibbons:2004js} as well as extreme Pomeransky-Sen'kov black rings \cite{Pomeransky}.

The metric \eqref{nhgeom} takes a Kaluza-Klein form.  There is an `internal' compact space $\Hcal$, parametrized by $(\phi^I,y^A)$, corresponding to a spatial cross-section of the black hole horizon.  More precisely, $\Hcal$ denotes a surface of constant $T$ and $R$ in \eqref{nhgeom}, with geometry
\begin{equation}
\label{Hmetric}
  d\hat{s}^2 = g_{IJ}(y) d\phi^I d\phi^J + g_{AB}(y) dy^A dy^B .
\end{equation}
Additionally, there is a non-compact $AdS_2$ space parametrized by $T$ and $R$.  Mixing between these two spaces is described by the terms $-k^I R dT$, which can be thought of as `Kaluza-Klein gauge fields' associated to a $U(1)^n$ gauge group. These preserve the symmetries of $AdS_2$ because the associated field strengths $k^I dT \wedge dR$ are proportional to the volume form of $AdS_2$; they describe homogeneous electric fields.

Our strategy will be to decompose perturbations as scalar fields in $AdS_2$, with the effective mass of these scalar fields given by eigenvalues of some operator on $\Hcal$.  This is more complicated than a standard (linearized) Kaluza-Klein reduction because the `KK gauge fields' are non-vanishing in the background geometry.  Fields with non-vanishing $\phi^I$ dependence will be charged with respect to the $AdS_2$ gauge fields.  In detail, the decomposition is as follows. 

\subsubsection{Scalar fields}
Consider first a complex scalar field $\Psi(T,R,\phi^I,y^A)$ satisfying the Klein-Gordon equation
\begin{equation}
 \left( \nabla^2 - M^2  \right) \Psi= 0.
\end{equation}
We start with a separable ansatz
\begin{equation}
  \Psi(T,R,\phi,y) = \chi_0(T,R) Y(\phi,y)
\end{equation}
and Fourier decompose $Y$ along the periodic directions $\phi^I$:
\begin{equation}
  Y(\phi,y) = e^{i m_I \phi^I} \Ybb(y) .
\end{equation}
The Klein-Gordon equation separates.  The function $\chi_0(T,R)$ satisfies the equation of a massive charged scalar field in $AdS_2$ with a homogeneous electric field.  More precisely, we write the $AdS_2$ metric and gauge field $A_2$ as
\begin{equation} \label{ads2}
  ds^2 = -R^2 dT^2 + \frac{dR^2}{R^2}, \qquad A_2 = -R \, dT,
\end{equation}
and introduce a gauge-covariant derivative for an $AdS_2$ scalar with charge $q$:
\begin{equation}\label{eqn:ads2deriv}
  D \equiv \nabla_2 - i q A_2,
\end{equation}
where $\nabla_2$ is the Levi-Civita connection associated with the $AdS_2$ metric.  
The scalar $\chi_0$ satisfies the equation of an $AdS_2$ scalar with charge $q$ and  squared mass $\mu^2 = \la + q^2$:
\begin{equation}
 \left( D^2 - \la - q^2\right) \chi_0= 0
\end{equation}
where the charge $q$ is given by\footnote{We are considering $AdS_2$ with a single gauge field $A= -R dT$. We could consider $AdS_2$ with multiple gauge fields, as is natural from the KK perspective, $A^I = -k^I R dT$. We would then obtain an $AdS_2$ scalar with charge $m_I$ with respect to $A^I$.  However, for fields of higher spin, it turns out to be more useful to consider a single gauge field.  The motivation for taking the separation constant to be $\la = \mu^2 - q^2$ rather than $\mu^2$ itself will also become apparent when we consider higher spin fields.}
\begin{equation}
 q = m_I k^I
\end{equation}
and the separation constant $\la$ is given by the eigenvalue equation
\begin{equation}\label{eqn:Oscalar}
 \Ocal{0} Y  \equiv  - \nablah_\mu\big( L(y)^2 \nablah^\mu Y\big) +L(y)^2  (M^2-q^2) Y= \la Y,
\end{equation}
where $\nablah$ is the Levi-Civita connection on $\Hcal$ and $\mu,\nu,\ldots$ denote indices on $\Hcal$, raised and lowered with the metric on $\Hcal$.

Note that $\la$ is real because $\Ocal{0}$ is self-adjoint with respect to the inner product
\begin{equation}
 (Y_1,Y_2) = \int_\Hcal \bar{Y}_1 Y_2\,\, \mathrm{d(vol)} .
\end{equation}
This self-adjointness also guarantees that the harmonics $Y$ form a complete set and hence any solution $\Psi$ can be expanded as a sum of separable solutions of the above form.\footnote{
Note also that $\Ocal{0}$ commutes with the Lie derivative with respect to $\pd/\pd{\phi^I}$ and hence eigenfunctions of $\Ocal{0}$ may be assumed to have the $\phi^I$ dependence assumed above.}

\subsubsection{Gravitational perturbations}
The same procedure works for the linearized gravitational field, described by $\Omega_{ij}$.  We give the details in Appendix \ref{app:nhframe} and summarize the results here.  We employ the gauge-invariant decoupled equation obtained in Ref.\ \cite{decoupling}, and start with a separable ansatz
\begin{equation}
 \Omega_{ij} = {\rm Re} \left[ \chi_2(T,R) Y_{ij} (\phi,y) \right].
\end{equation}
Since we are choosing our null basis vector $\ell$ and $n$ to be tangent to the null geodesic congruences with vanishing expansion, rotation and shear, i.e., to $-RdT \pm dR/R$, it follows that the spatial basis vectors $m_i$ span $\Hcal$. Therefore, we can regard $Y_{ij}$ as the components of a symmetric traceless tensor $Y_{\mu\nu}$ on $\Hcal$.  We take a Fourier decomposition of this tensor, that is we assume that
\begin{equation}
 \Lcal_I Y_{\mu\nu} = i m_I Y_{\mu\nu},
\end{equation}
where $\Lcal_I$ is the Lie derivative with respect to $\pd/\pd \phi^I$.  We can again perform a separation of the perturbation equation for $\Omega_{ij}$, and show that it reduces to the equation of a massive charged scalar in $AdS_2$, satisfying
\begin{equation}
  \left( D^2 - q^2- \lambda \right) \chi_2 = 0.
\end{equation}
where $D$ was defined in (\ref{eqn:ads2deriv}), the charge is given by
\begin{equation}
 q = m_I k^I + 2i
\end{equation}
and the separation constant $\la$ by the eigenvalue equation
\begin{equation}
 (\Ocal{2} Y)_{\mu\nu} = \la Y_{\mu\nu}
\end{equation}
for an operator
\begin{multline}
  (\Ocal{2} Y)_{\mu\nu}
   =  -\frac{1}{L^4}\nablah^\rho \left(L^6 \nablah_\rho Y_{\mu\nu}\right)
      + \left(6-(k^I m_I)^2 - \tfrac{4}{L^2} k_\rho k^\rho - 2(d-4)\La L^2 \right)Y_{\mu\nu} \\
      + 2L^2\left(\hat{R}_{(\mu|\rho} 
             + \hat{R}\hat{g}_{(\mu|\rho}\right)Y^\rho_{\phantom{\rho} |\nu)}
      - 2L^2\hat{R}_{\mu\phantom{\rho}\nu}^{\phantom{\mu}\rho\phantom{\nu}\sig}
                                          Y_{\rho\sig} \\
   + \Big[ - (dk)_{(\mu|\rho} - \tfrac{2}{L^2}\left(d(L^2)\wedge k\right)_{(\mu|\rho}\\
            + 2 \left( k-d(L^2)\right)_{(\mu|}\nablah_\rho 
            - 2 \left( k-d(L^2)\right)_{\rho} \nablah_{(\mu|}
            \Big]
                        Y^\rho_{\phantom{\rho} |\nu)}.
 \label{eqn:Ograv}
\end{multline}
In this expression, $\hat{R}_{\mu\nu\rho\sigma}$ is the Riemann tensor on ${\cal H}$ (with $\hat{R}_{\mu\nu}$ and $\hat{R}$ the Ricci tensor and Ricci scalar), indices are raised and lowered with the metric (\ref{Hmetric}) on ${\cal H}$, $k$ is the Killing vector field on ${\cal H}$ defined by
\begin{equation}
 k = k^I \frac{\partial}{\partial \phi^I}
\end{equation}
and $(dk)_{\mu\nu} \equiv 2 \hat{\nabla}_{[\mu} k_{\nu]}$. We have written $\Ocal{2}$ in a covariant way, so that it can be evaluated without having to use the particular coordinates on $\Hcal$ that we introduced above. The explicit $m_I$ dependence enters only via $k^I m_I$, which can be determined from
\begin{equation}
\label{kdotm}
 {\cal L}_k Y_{\mu\nu} = i k^I m_I Y_{\mu\nu} .
\end{equation}
We define an inner product between traceless, symmetric, valence 2 tensors on $\Hcal$ by
\begin{equation}\label{eqn:gravinner}
  (Y_1,Y_2) \equiv \int_\Hcal L^4 \bar{Y}_1^{\mu\nu} Y_{2\mu\nu} \; \mathrm{d(vol)}.
\end{equation}
It can be shown that $\Ocal{2}$ is self-adjoint with respect to this inner product. This implies  that the eigenvalues $\la$ are real.

The function $\chi_2(T,R)$ satisfies the equation of a charged scalar in $AdS_2$ where the mass $\mu$ is given by
\begin{equation}
 \mu^2 = q^2 + \la .
\end{equation}
Note that $q$ is {\it complex}.  This has been observed previously for gravitational perturbations of the NHEK geometry \cite{Amsel:2009ev,Dias:2009ex}. Self-adjointness implies that $\la$ is real and hence $\mu^2$ also is complex but the combination $\mu^2 - q^2$ is always real.

We should mention that the use of the gauge-invariant quantity $\Omega_{ij}$ to describe metric perturbations implies that we will not be able to study certain non-generic perturbations that preserve the algebraically special property of the background geometry and hence have $\Omega_{ij} = 0$. In particular, we will miss perturbations that deform the near-horizon geometry into another near-horizon geometry.\footnote{
This is analogous to the fact that the formalism of Ref.\ \cite{Teukolsky} misses modes corresponding to infinitesimal changes in the mass or angular momentum of the Kerr solution.}

\subsubsection{Electromagnetic Perturbations}

Finally, we can also analyse the behaviour of Maxwell fields. In a Kundt background, these satisfy a decoupled equation in terms of a quantity $\vphi_i = F_{\mu\nu} \ell^\mu m_i^\nu$ \cite{decoupling}.  Similarly to previous cases, we write
\begin{equation}
  \vphi_i (T,R,\phi^I,y^A) = \mathrm{Re} \big[ \chi_1 (T,R) Y_i(\phi^I,y^A) \big] .
\end{equation}
The decoupled equation for $\vphi_i$ can be separated to give the equation of a charged scalar in $AdS_2$:
\begin{equation}
  (D^2-\la-q^2)\chi_1 = 0
\end{equation}
where the charge is
\begin{equation}
 q =  k^I m_I + i,
\end{equation}
the mass $\mu$ is given by $\mu^2 = q^2 + \lambda$, and $\lambda$ is given by
\begin{equation}
  (\Ocal{1} Y)_\mu = \la Y_\mu 
\end{equation}
where
\begin{multline}
  (\Ocal{1} Y)_\mu 
   =  -\frac{1}{L^2}\nablah^\rho \left(L^4 \nablah_\rho Y_\mu \right)
      + \left(2-(k^I m_I)^2 - \tfrac{5}{4L^2} k_\nu k^\nu - \tfrac{d-6}{2}\La L^2 \right)Y_\mu \\
      + L^2(\hat{R}_{\mu\nu} + \tfrac{1}{2} \hat{R}\hat{g}_{\mu\nu})Y^\nu
   + \left( - \tfrac{1}{2} (dk)_{\mu\nu} 
            + 2 \left( k-d(L^2)\right)_{[\mu} \nablah_{\nu]}
            - \tfrac{1}{L^2} (dL^2)_{[\mu} k_{\nu]} 
     \right) Y^\nu .
 \label{eqn:Omax}
\end{multline}
This is again self-adjoint, this time with respect to the inner product
\begin{equation}\label{eqn:maxinner}
  (Y_1,Y_2) \equiv \int_\Hcal L^2 \bar{Y}_1^{\mu} Y_{2\mu} \; \mathrm{d(vol)}.
\end{equation}
Just as in the gravitational case, our use of the quantity $\vphi_i$ to describe the Maxwell field means that we miss certain non-generic perturbations with $\vphi_i = 0$. For Myers-Perry black holes, we will miss the perturbation corresponding to turning on electric charge in the background spacetime. 

\subsection{Behaviour of solutions}\label{sec:behaviour}

We've seen that for a scalar field, linearized gravitational field, or Maxwell field, we can reduce the equation of motion to that of a massive, charged, scalar field $\chi_b(T,R)$ in $AdS_2$ with a homogeneous electric field (\ref{ads2}). Solutions of this equation of motion were considered in Refs.\ \cite{Strominger:1998yg,Amsel:2009ev,Dias:2009ex}. At large $R$, they behave as $\chi_b \sim R^{-\Delta_{\pm}}$ where
\begin{equation}\label{eqn:modefreq}
  \Delta_\pm = \frac{1}{2} \pm \sqrt{ \mu^2 - q^2 + \frac{1}{4} } .
\end{equation}
Therefore solutions grow or decay as real powers of $R$ if the `effective BF bound' is respected:
\begin{equation}\label{eqn:ads2stabbound}
  \mu^2 - q^2 \ge -\frac{1}{4}.
\end{equation}
If this bound is violated then solutions oscillate at infinity.  In the uncharged case ($q=0$), it is known that boundary conditions can be imposed that lead to stable, causal, dynamics when the bound is respected \cite{BF,Ishibashi:2004wx}. If the bound is violated then no choice of boundary conditions leads to stable, causal, dynamics \cite{Ishibashi:2004wx}. Motivated by this, we make the following definition for the remainder of the paper:

\medskip
\noindent {\bf Definition}. {\it 
  A near-horizon geometry is \emph{unstable} against linearized gravitational (or scalar field or Maxwell) perturbations if expanding in harmonics on $\Hcal$ gives a massive, charged, scalar field in $AdS_2$ that violates the bound (\ref{eqn:ads2stabbound}).}
\medskip

This is just introducing some terminology, we are not claiming anything about the dynamics of a scalar field in $AdS_2$ when (\ref{eqn:ads2stabbound}) is violated. Of course, it would be interesting to see if the arguments of Ref.\ \cite{Ishibashi:2004wx} could be extended to the charged case to show that violation of (\ref{eqn:ads2stabbound}) implies that there exists no choice of boundary conditions for which the scalar field has stable dynamics. However, such considerations are not relevant to this paper, as we are interested in the question of whether violation of (\ref{eqn:ads2stabbound}) implies instability of the full black hole geometry rather than just its near-horizon geometry. In fact, the results of Refs.\ \cite{Amsel:2009ev,Dias:2009ex} show that it probably doesn't make sense to consider perturbations of the near-horizon geometry as a spacetime in its own right since there will be a large backreaction when one goes beyond linearized theory.

We showed above that $\mu^2-q^2 = \lambda$, the eigenvalue of a self-adjoint operator $\Ocal{b}$. Hence, our condition for instability of the near-horizon geometry is the existence of an eigenvalue $\lambda<-1/4$. So the question of stability reduces to studying the spectrum of these operators on ${\cal H}$.  In the next section we shall study the spectrum of these operators for the case of extreme cohomogeneity-1 MP black holes. 

\section{Cohomogeneity-1 extreme Myers-Perry black holes} \label{sec:nhmp}

\subsection{Metric and near-horizon limit}

We shall now illustrate the methods described above with an example.  Consider a Myers-Perry-(AdS) black hole \cite{mp,Hawking:1998kw,Gibbons:2004js} in odd dimension $d=2N+3$, with all angular momentum parameters set to be equal, $a_I=a$. Such a black hole has enhanced rotational symmetry; the $U(1)^{N+1}$ is enlarged to $U(N+1)$, i.e.\ the symmetry is that of a homogeneously squashed $S^{d-2} = S^{2N+1}$. The metric is cohomogeneity-1, i.e., it depends non-trivially on a single coordinate. This makes the study of gravitational perturbations of this class of black holes more tractable than the general case, and certain types of perturbation of the full black hole geometry have been studied previously \cite{Kunduri:2006,Murata:2008,Dias:2010eu}.

The metric for the full black hole solution can be written in the form \cite{Kunduri:2006}
\begin{equation}\label{eqn:mpmetric}
  ds^2 = -\frac{V(r)}{h(r)^2} dv^2 +  \frac{2drdv}{h(r)} + r^2 h(r)^2 (d\hat{\psi} + \Acal - \Om(r)dv)^2 
         + r^2\hat{g}_{\al\beta} dx^\al dx^\beta
\end{equation}
where $(v,r,\hat{\psi},x^\al)$ are ingoing Eddington-Finkelstein type coordinates, $\hat{\psi}$ has period $2\pi$,
\begin{equation}
  V(r) = 1 + \frac{r^2}{l^2} 
         + \left(\frac{r_0}{r}\right)^{2N} \left(-1 + \frac{a^2}{l^2} + \frac{a^2}{r^2}
           \right),
\end{equation}
\begin{equation}
  h(r) = \sqrt{1 + \frac{a^2}{r^2}\left(\frac{r_0}{r}\right)^{2N}}\eqand
  \Om(r) = \frac{a}{r^2 h(r)^2}\left(\frac{r_0}{r}\right)^{2N}.
\end{equation}
The solution is parameterized by three quantities with the dimensions of length: $r_0$, $a$ (which determines the ratio of angular momentum to mass), and $l$ (the AdS radius). We are writing the $S^{2N+1}$ as a $U(1)$ fibration over $\CP{N}$, with $\hat{g}_{\al\beta}$ the Fubini-Study metric on $\CP{N}$ (normalized to have Ricci tensor $2(N+1) \hat{g}_{\al\beta}$) and $\Acal = \Acal_\al dx^\al$ satisfying $d\Acal = 2\Jcal$, where $\Jcal$ is the K\"ahler form on $\CP{N}$.  The metric satisfies the vacuum Einstein equation
\begin{equation}\label{eqn:einstein}
  R_{ab} = -\frac{d-1}{l^2} g_{ab} \equiv \La g_{ab},
\end{equation}
and is asymptotically $AdS_d$ with radius $l$. The limit $l\rightarrow \infty$ gives the asymptotically flat MP solution.

The event horizon lies at $r=r_+$, with $V(r_+)=0$.  This family of black holes admits an extremal limit, i.e. there exists a value of $a$ for which $V'(r_+)=0$.  In this case, the solution is uniquely labelled by $l$ and $r_+$, with
\begin{equation}
  r_0^N = r_+^{N+2} \sqrt{N+1} \left(\frac{1}{r_+^2} + \frac{1}{l^2} \right), \qquad
  a^2   = \frac{r_+^2 l^2}{N+1} \left(\frac{(N+1)r_+^2 + Nl^2}{(r_+^2+l^2)^2}\right). 
\end{equation}
To obtain the near-horizon limit, we define new coordinates $\tilde{r},\tilde{v}, \tilde{\psi}$ by 
\begin{equation}
  r = r_+ + \eps \tilde{r}, \quad v = \frac{\tilde{v}}{\eps} \eqand \hat{\psi} = \tilde{\psi} + \Om(r_+) v,
\end{equation}
and then take the limit $\eps\rightarrow 0$, to obtain a metric
\begin{equation}\label{eqn:ds2exact}
  ds^2 =  -\frac{V''(r_+)\tilde{r}^2}{2h(r_+)^2} d\tilde{v}^2 + \frac{2d\tilde{r}d\tilde{v}}{h(r_+)}
         + r_+^2 h(r_+)^2 \left(d\tilde{\psi} + \Acal - \Om'(r_+)\tilde{r}d\tilde{v}\right)^2 
         + r_+^2\hat{g}_{\al\beta} dx^\al dx^\beta.
\end{equation}
Finally, to simplify this, and recover a form of the metric more similar to that used in the discussion above, we define new coordinates $(T,R,\psi,x^\al)$ by
\begin{equation}
  T = \frac{V''(r_+)}{2h(r_+)} \tilde{v} + \frac{1}{\tilde{r}},\qquad
  R = \tilde{r}, \qquad
  \psi = \tilde{\psi} - \frac{2h(r_+) \Om'(r_+)}{V''(r_+)}\log(\tilde{r})
\end{equation}
and define constants
\begin{eqnarray}
\label{Ldef}
 \frac{1}{L^2} &=& \frac{V''(r_+)}{2} 
                = 2(N+1) \left( \frac{N}{r_+^2} + \frac{N+2}{l^2} \right)\label{eqn:Ldef} \\ 
  B^2 &=& r_+^2 h(r_+)^2 
       = (N+1) r_+^2 \left(1+ \frac{r_+^2}{l^2}\right), \label{eqn:Bdef}\\
  \Om &=& \frac{2h(r_+) \Om'(r_+)}{V''(r_+)} 
       =  \frac{-1}{(N+1)\big(N+(N+2)(r_+/l)^2\big)}\sqrt{\frac{Nl^2 + (N+1)r_+^2}{l^2+r_+^2}} , \\
  \frac{1}{E}  &=& \frac{B\Om}{2L^2} 
                  = -\Big(1+\frac{r_+^2}{l^2}\Big) 
                    \sqrt{(N+1)\left(\tfrac{N+1}{l^2} + \tfrac{N}{r_+^2}\right)} \label{eqn:Edef}.
\end{eqnarray}
This gives a simple form for the near-horizon metric:
\begin{equation}\label{eqn:metric}
  ds^2 = L^2(-R^2 dT^2 + \frac{dR^2}{R^2}) + B^2 \left(d\psi + \Acal - \Om R dT\right)^2 
         + r_+^2 \hat{g}_{\al\beta} dx^\al dx^\beta.
\end{equation}
As expected, this metric takes the form of a $(d-2)$-dimensional manifold $\Hcal$ fibred over $AdS_2$.  Here, $\Hcal$ is a homogeneously squashed $(d-2)$-sphere, with metric
\begin{equation}
  ds_{d-2}^2 = B^2 \left(d\psi + \Acal\right)^2 + r_+^2 \hat{g}_{\al\beta} dx^\al dx^\beta,
\end{equation}
where $\hat{g}$ is the metric on $\CP{N}$ as above and $\psi$ has period $2\pi$.

We are writing the metric in a form that makes manifest its enhanced symmetry, rather than in the form \eqref{nhgeom} (which makes manifest only the Killing directions $\pd/\pd\phi^I$).  Since we know that the near-horizon geometry of a general extreme MP solution \emph{can} be written in the form \eqref{nhgeom} \cite{Figueras:2008qh} it follows that there must be a coordinate transformation that would allow us to bring our metric to this form.  However, it is not necessary to perform such a transformation since the operators $\Ocal{b}$ on $\Hcal$ are defined in a covariant way.  We can read off the vector $k$ by looking at the cross-terms proportional to $\pd/\pd T$ in the inverse metric:
\begin{equation}
  \frac{1}{2L^2} \left(-2\Om \pard{\psi} \pard{T} \right) = 
  \frac{1}{2L^2} \left( -2k^I \pard{\phi^I} \pard{T} \right)
\end{equation}
and hence
\begin{equation}\label{eqn:nhmpk}
  k = \Om \left(\pard{\psi}\right).
\end{equation}
We can Fourier decompose our perturbation in the $\psi$ direction, i.e. assume dependence $e^{im\psi}$ so that eigenfunctions $Y$ on $\Hcal$ obey $\Lcal_k Y = i\Om m Y$.  Equation \eqref{kdotm} now enables us to read off 
\begin{equation}
  k^I m_I = \Omega m .
\end{equation}
For these black holes, the condition (\ref{eqn:axicondition}) reduces to $m=0$.  However, we will obtain results for general $m$.  We will determine the spectrum of our operators $\Ocal{b}$ by expanding them in harmonics on $\CP{N}$, with metric $\gh_{\al\beta}$ (where $\al,\beta,\ldots$ are indices on $\CP{N}$, raised and lowered with $\gh$). From the $\CP{N}$ perspective, $m$ acts like a charge which couples to the `gauge field' $\Acal$ (see \cite{Kunduri:2006}).  We therefore define a charged covariant derivative on $\CP{N}$
\begin{equation}\label{eqn:Dhatdef}
  \Dcalh_\al = \hat{D}_\al - im \Acalh_\al
\end{equation}
where $\hat{D}$ is the Levi-Civita connection on $\CP{N}$.
%

\subsection{Scalar field perturbations}\label{sec:scalarfield}

As a simple first example, we show how to deal with massive scalar field perturbations. The operator $\Ocal{0}$ defined by (\ref{eqn:Oscalar}) reduces to
\begin{equation}\label{eqn:massivescalarperts}
   \Ocal{0} Y =  -\frac{2Nm^2L^4}{r_+^4}Y - \frac{L^2}{r_+^2} \Dcalh^2 Y + L^2M^2 Y,
\end{equation}
acting on functions $Y(\psi,x) = e^{im\psi} \Ybb(x)$.  We shall assume that the $AdS_d$ BF bound is respected, i.e. that
\begin{equation} \label{eqn:adsdbf}
  M^2 \geq -\frac{(d-1)^2}{4l^2} = -\frac{(N+1)^2}{l^2}.
\end{equation}
Scalar eigenfunctions of the charged covariant Laplacian $\Dcalh^2$ on $\CP{N}$ were studied in \cite{Hoxha:2000}.  For each integer $m$, there exist $\CP{N}$ scalars $\Ybb(x)$ satisfying 
\begin{equation} \label{eqn:scalarYbb}
  (\Dcalh^2 + \lak{S}) \Ybb = 0,
\end{equation}
for eigenvalues
\begin{equation}\label{eqn:scalarevals}
  \lak{S} = 4\kap(\kap+N) + 2|m|(2\kap+N) \qquad \kap = 0,1,2,\dots.
\end{equation}
Hence, the eigenvalues of $\Ocal{0}$ are
\begin{equation}
  \la = \frac{\left(4\kap(\kap+N) + 2|m|(2\kap+N)\right)L^2}{r_+^2} -\frac{2Nm^2L^4}{r_+^4} + M^2L^2.
\end{equation}
Therefore, for large $|m|$, $\la$ becomes arbitrarily negative and the BF bound \eqref{eqn:ads2stabbound} is always violated.  However, for the axisymmetric modes $m=0$ that are relevant for our conjecture, the eigenvalues are given by 
\begin{equation}
  \frac{\la}{L^2} = \frac{4\kap(\kap+N)}{r_+^2} + M^2 
\end{equation}
for non-negative integers $\kap$ (recall that $L$ is defined by (\ref{Ldef})). 

Consider first asymptotically flat black holes $l\rightarrow \infty$ and $M^2 \ge 0$.  Here, we manifestly have $\la \geq 0$, and hence the $AdS_2$ BF bound is not violated.

This is not always the case for asymptotically $AdS$ black holes. Clearly there is no problem if $M^2 \ge 0$. However, if $M^2<0$ then it is possible for the $AdS_2$ BF bound to be violated even if the $AdS_d$ BF bound is respected \cite{Dias:2010ma}. Consider for example the case in which the $AdS_d$ bound is saturated. Then a mode labelled by $\kap$ violates the $AdS_2$ BF bound if
\begin{equation}
  \frac{r_+^2}{l^2} > \frac{8\kap(\kap+N) }{N(N+1)}+1,
\end{equation}
that is, for sufficiently large black holes.  In this case, our conjecture predicts that the scalar field should be unstable in the full black hole geometry. This issue was investigated numerically in Ref. \cite{Dias:2010ma}.  It was found that the full black hole is indeed unstable, and there exists a new nonlinear family of `hairy' rotating black holes. In Section~\ref{sec:nhinstab} we shall prove analytically that the full black hole solution must be unstable.

\subsection{Gravitational perturbations of asymptotically flat BHs}
                                                                                
We now consider the more complicated case of gravitational perturbations.  The calculations here are significantly more involved.  In this section we will merely give the results for different classes of perturbation mode, reserving the details of the calculations for Appendix \ref{app:technical}.

Our approach to determining the eigenvectors $Y_{\mu\nu}$ of $\Ocal{2}$ is to decompose $Y_{\mu\nu}$ into parts parallel and perpendicular to $\CP{N}$ and then expand each part in terms of harmonics on $\CP{N}$, assuming dependence $e^{im\psi}$ along the $S^1$ fibre. By `harmonics', we mean eigenfunctions of the charged $\CP{N}$ Laplacian $\hat{{\cal D}}^2$. They can be divided into scalar, vector, and (traceless) tensor types \cite{Kunduri:2006,Martin:2008pf,Dias:2010eu} where vector and tensor harmonics are transverse with respect to the derivatives $\Dcalh_\al$ and $\Jcalh_\al{}^\beta \Dcalh_\beta$. See Ref. \cite{Martin:2008pf} for detailed discussion of this decomposition. 
The orthogonality properties of these different types of harmonic implies that eigenfunctions of $\Ocal{2}$ must each be built from $\CP{N}$ harmonics of a particular type (scalar, vector or tensor) and with the same eigenvalue of $\Dcalh^2$.

The modes that are relevant to our conjecture are those that are $\psi$ independent, i.e.\ those with $m=0$.  Therefore, we only list our results in this case, although in Appendix \ref{app:technical} we derive all of these results for general $m$.  It turns out that, as in the scalar field case, the coefficient of $m^2$ in these eigenvalues is always negative, and hence for sufficiently large $|m|$ there are instabilities in every sector of perturbations of the near-horizon geometry.

We begin with the asymptotically flat case, corresponding to $l\rightarrow \infty$.  

\subsubsection{Tensor modes}\label{sec:tensorevals}

These eigenfunctions $Y_{\mu\nu}$ have components only in the direction of $\CP{N}$, and are proportional to a transverse, traceless, tensor harmonic on $\CP{N}$. Such harmonics exist only for $N>1$ ($d>5$). Tensor perturbations of the full black hole geometry were considered in Ref. \cite{Kunduri:2006}.  In the asymptotically flat case, no evidence of any instability was found near extremality. Hence, if our conjecture holds, we would not expect to find any unstable modes satisfying \eqref{eqn:axicondition} (i.e. $m=0$) in this sector.

We find that the eigenvalues $\la$ of $\Ocal{2}$ are given by 
\begin{equation}\label{eqn:tensorevalsflat}
  \la = \frac{2\kap(\kap+N) + 2N(1-\sig)}{N(N+1)},
\end{equation}
where $\kap=0,1,2,\ldots$, and the parameter $\sig = \mp 1$ separates two different classes of tensor harmonic which are respectively Hermitian, or anti-Hermitian, on $\CP{N}$ (more details are given in Section \ref{sec:TTmodes}).

The eigenvalues $\lambda$ are manifestly non-negative.  Hence the effective BF bound $\lambda \ge -1/4$ is respected and there is no instability of the near-horizon geometry in this sector.  Hence our conjecture is consistent with the results of Ref. \cite{Kunduri:2006}.

\subsubsection{Vector modes}

Next, we move on to study vector-type perturbation modes. Again, these exist only for $N>1$ ($d>5$). These have not been previously studied in the literature, so we have no numerical results for the full black hole geometry to compare our results to. 

For vector-type perturbations $Y_{\mu\nu}$ is written as a linear combination of three different types of term built from a $\CP{N}$ vector harmonic and its derivatives, so it is determined by the three coefficients in this expansion. Acting with $\Ocal{2}$ has the same effect as acting with a certain $3\times 3$ matrix on these coefficients. Hence finding the eigenvalues of $\Ocal{2}$ for vector type perturbations reduces to finding the eigenvalues of a $3\times 3$ matrix. The elements of this matrix involve the eigenvalue of the vector harmonic on $\CP{N}$, which is labelled by a non-negative integer $\kap$ (and the integer $m$). Perhaps surprisingly, the eigenvalues of $\Ocal{2}$ turn out to be rational (given here for $m=0$):
\begin{equation}\label{eqn:gravvecevals}
  \la = \frac{2(N+(\kap+1)^2)}{N(N+1)}, \quad
  \frac{2(\kap+2)(\kap+N+1)}{N(N+1)}, \quad
  \frac{2\left(N^2 + (\kap+2)^2 + N(2\kap+5)\right)}{N(N+1)}.
\end{equation}
These are all manifestly positive, so there is no violation of the generalized $AdS_2$ BF bound in this sector.

\subsubsection{Scalar modes}\label{sec:gravscalarevalsflat}

The most complicated sector is that of scalar type gravitational perturbations.  In this case, $Y_{\mu\nu}$ is written as a linear combination of six terms, each of which is constructed from $\CP{N}$ scalar harmonics and their derivatives. 
Harmonics are again labelled by an integer $\kappa \geq 0$, as well as $m$. Acting with $\Ocal{2}$ has the effect of acting with a $6 \times 6$ matrix. Hence determining the eigenvalues of $\Ocal{2}$ is equivalent to determining the eigenvalues of a $6 \times 6$ matrix. For the special cases $\kappa=0,1$ some combinations of derivatives of the $\CP{N}$ harmonics vanish, which a corresponding reduction in the size of the matrix. There is also a reduction in size for the special case of $N=1$ (i.e. $d=5$) for which the matrix is generically $5 \times 5$. In all cases, we again find that the eigenvalues of $\Ocal{2}$ are rational. 

For $\kap=0$, there is just one eigenvalue
\begin{equation}\label{eqn:gravkap0}
  \la = \frac{2(2N+1)}{N} 
\end{equation}
which is manifestly positive, and hence there is no instability here.

For $\kap=1$, the eigenvalues $\la$ correspond to the eigenvalues of a $4\times 4$ matrix ($3 \times 3$ for $N=1$). They are
\begin{equation}
  \la  = \frac{2}{N}, \quad \frac{2(N+1)}{N}, \quad \frac{2(N+2)}{N},\quad \frac{4(N+2)}{N},
\end{equation}
which are again all positive. The second eigenvalue listed above is absent for $N=1$.

Things get more interesting for $\kap\geq 2$, where we have a $6 \times 6$ matrix ($5\times 5$ for $N=1$).  The eigenvalues are given by
\begin{multline}\label{eqn:gravscalarevecs}
  \la = \frac{2(\kap-1)(\kap-N-1)}{N(N+1)}, \quad
                     \frac{2\kap(\kap-1)}{N(N+1)}, \quad 
                   \frac{2\kap(\kap+N)}{N(N+1)}, \\
                   2+\frac{2\kap(\kap+N)}{N(N+1)}, \quad
                   \frac{2(\kap+N)(\kap+N+1)}{N(N+1)},\quad
                   \frac{2(\kap+1+N)(\kap+2N+1)}{N(N+1)}.
\end{multline}
The fourth of these is absent for $N=1$.
Five of these eigenvalues are manifestly non-negative, so in order to check for an instability of the near-horizon geometry, we need only to analyse whether there exist $\kap$, $N$ such that
\begin{equation}\label{eqn:minevalflat}
  \frac{2(\kap-1)(\kap-N-1)}{N(N+1)} < -\frac{1}{4}.
\end{equation}
We list the values of the left hand side explicitly in Table \ref{tab:evals1} for all $\kap=2,\dots 10$, in dimensions $d=5,7,\dots,23$. 

\begin{table}[htb]
\begin{center}
  \begin{tabular}{|c|c||c|c|c|c|c|c|c|c|c|}
\hline
       &   & $\kap$&       &       &       &        &   &   &   &  \\
   $d$ &$N$&     2 &     3 &     4 &     5 &     6  & 7 & 8 & 9 & 10\\ \hline\hline
     5 & 1 &  0.00 & 2.00  &  6.00 & 12.00 & 20.00 & 30.00 & 42.00 & 56.00 & 72.00\\
     7 & 2 & \bf-0.33 & 0.00  &  1.00 & 2.67  & 5.00  & 8.00  & 11.70 & 16.00 & 21.00\\
     9 & 3 & \bf-0.33 & \bf-0.33 &  0.00 & 0.67 & 1.67  & 3.00  & 4.67  & 6.67 & 9.00\\
    11 & 4 & \bf-0.30 & \bf-0.40 & \bf-0.30 & 0.00  & 0.50  & 1.20  & 2.10  & 3.20 & 4.50\\
    13 & 5 & \bf-0.27 & \bf-0.40 & \bf-0.40 & \bf-0.27 & 0.00  & 0.40  & 0.93  & 1.60 & 2.40\\
    15 & 6 & -0.24 & \bf-0.38 & \bf-0.43 & \bf-0.38 & -0.24 & 0.00  & 0.33  & 0.76 & 1.29\\
    17 & 7 & -0.21 & \bf-0.36 & \bf-0.43 & \bf-0.43 & \bf-0.36 & -0.21 & 0.00  & 0.29 & 0.64\\
    19 & 8 & -0.19 & \bf-0.33 & \bf-0.42 & \bf-0.44 & \bf-0.42 & \bf-0.33 & -0.19 & 0.00 & 0.25\\
    21 & 9 & -0.18 & \bf-0.31 & \bf-0.40 & \bf-0.44 & \bf-0.44 & \bf-0.40 & \bf-0.31 & -0.18& 0.00\\
    23 &10 & -0.16 & \bf-0.29 & \bf-0.38 & \bf-0.44 & \bf-0.46 & \bf-0.44 & \bf-0.38 & \bf-0.29& -0.16\\\hline
  \end{tabular}
  {\it\small\caption{\label{tab:evals1} Smallest eigenvalue of $\Ocal{2}$ for $m=0$, in the case of asymptotically flat extremal cohomogeneity-1 Myers-Perry black holes in dimensions $d=5,7,\dots 23$, for modes $\kap=2,\dots 10$.  The BF bound is $-1/4$, eigenvalues violating this bound, and indicating an instability of the near-horizon geometry, are shown in bold (NB: all of these values are rational numbers determined by \eqref{eqn:gravscalarevecs}, we give decimal approximations here for readability purposes.)}}
\end{center}
\end{table}

In dimension $d=5$ there are no modes that violate the effective BF bound, and we conclude that there are no unstable scalar modes of the near-horizon geometry that satisfy the condition (\ref{eqn:axicondition}). Therefore we do not predict any instability of the full black hole in this case. This is consistent with a study of linearized perturbations of the full black hole \cite{Murata:2008}, which did not find any evidence of instability near extremality.

Our main result in this section is that for $d \ge 7$ there is always at least one mode that violates the effective BF bound and hence the near-horizon geometry is unstable. Since this mode respects (\ref{eqn:axicondition}), our conjecture predicts that the full black hole solutions should be unstable. Perturbations of the full non-extreme black hole were studied in Ref. \cite{Dias:2010eu}. For $d = 9$ it was found that $\kap=2$ scalar perturbations are unstable near extremality, in agreement with our conjecture. However no instability was found for the cases $d=7$, $\kap=2$ or $d=9$, $\kap=3$ for which we predict that one should be present.

The reason for this discrepancy is that the results of Ref.~\cite{Dias:2010eu} do not get close enough to extremality to see the instability that we predict.  At our request, J.E.~Santos has kindly repeated the numerical analysis of Ref.~\cite{Dias:2010eu} for black holes that are very close to extremality. He finds instabilities that were missed in the analysis of Ref.~\cite{Dias:2010eu}.  Let $a_{\rm ext}$ denotes the value of $a$ at which the black hole becomes extreme. Table \ref{tab:jorge} gives the critical value of $1-a/a_{\rm ext}$ below which the black hole is unstable.\footnote{
There is no instability of the black hole for $\kap=1$ but Ref.~\cite{Dias:2010eu} showed that there is an instability of the corresponding black {\it string} close to extremality. For completeness, we give Santos' results for the critical values of $1-a/a_{\mathrm{ext}}$ for this instability: $ 4.116 \times 10^{-2}$, $ 8.347 \times 10^{-2}$, $ 1.351 \times 10^{-1}$, $ 1.517 \times 10^{-1}$ for $d=7,9,13,15$ respectively.} There is indeed an instability very near extremality for $d=7$, $\kap=2$ and $d=9$, $\kap=3$, for $d=13$ with $\kap = 2,3,4,5$ and for $d=15$ with $\kap = 3,4,5$, all in perfect agreement with our conjecture.\footnote{Santos tells us that he had difficulty getting good numerical accuracy for $d=11$.} He also finds that there are cases for which we do not predict an instability but nevertheless one exists (e.g. $d=7$, $\kap=3$), which emphasizes that our conjecture supplies a sufficient, but not necessary, condition for instability. 

\begin{table}
\centering
\small
\begin{tabular}{|c||c|c|c|c|c|c|}
\hline
               & $\kappa = 2$ & $\kappa = 3$ & $\kappa = 4$ & $\kappa = 5$ & $\kappa = 6$ & $\kappa = 7$ \\
\hline\hline
$ d = 7$       & $ 2.339 \times 10^{-5}$  & $ 2.507 \times 10^{-7}$  &  &  &  & \\
\hline
$ d = 9$       & $ 2.116 \times 10^{-3}$  & $ 2.942 \times 10^{-7}$  & $ 8.021 \times 10^{-9}$  &  &  & \\
\hline
$ d = 13$      & $ 1.504 \times 10^{-2}$  & $ 1.358 \times 10^{-3}$  & $ 2.112 \times 10^{-5}$  
               & $ 1.056 \times 10^{-6}$  & $ 7.35 \times 10^{-8}$  & \\
\hline
$ d = 15$      & $ 2.232 \times 10^{-2}$  & $ 3.463 \times 10^{-3}$  & $ 2.868 \times 10^{-4}$  
               & $ 5.05 \times 10^{-6}$  & $ 7.57 \times 10^{-7}$  & $ 6.103 \times 10^{-8}$  \\
\hline
\end{tabular}
{\it\small\caption{\label{tab:jorge}
Critical values $1-a/a_{\mathrm{ext}}$ below which an asymptotically flat, cohomogeneity-1, Myers-Perry black hole becomes unstable against scalar-type gravitational perturbations with the given $\kappa$.  
We are very grateful to Jorge Santos for permitting us to include these results.}}
\end{table}
In general dimension $d=2N+3$, straightforward algebra shows that a violation of the effective BF bound occurs if
\begin{equation}
  1 + \tfrac{N}{2} - \tfrac{1}{2} \sqrt{\tfrac{N(N-1)}{2}} < \kap 
          < 1 + \tfrac{N}{2} + \tfrac{1}{2} \sqrt{\tfrac{N(N-1)}{2}}. 
\end{equation}
This proves that for any $N\geq 2$, there is at least one integer value of $\kap$ for which the effective BF bound is violated.

\subsection{Gravitational perturbations of asymptotically $AdS$ BHs}

\subsubsection{Introduction}

We now move on to consider gravitational perturbations of cohomogeneity-1 Myers-Perry-AdS black holes. Ref.~\cite{Kunduri:2006} demonstrated that such black holes suffer a `superradiant' instability near extremality.  This instability corresponds to perturbations with $m\neq 0$, which are excluded from the scope of our conjecture. We shall consider eigenfunctions of $\Ocal{2}$ with $m=0$ to see if any new instability appears. Once again, we consider separately eigenfunctions of $\Ocal{2}$ constructed from tensor, vector and scalar harmonics on $\CP{N}$.

\subsubsection{Tensor modes}

The eigenvalues $\la$ of $\Ocal{2}$ are given by 
\begin{equation}\label{eqn:tensorevalsads}
  \frac{\la}{L^2}
    = 4(1-\sig) \left(\frac{N}{r_+^2}+\frac{N+1}{l^2}\right) + \frac{4\kap(\kap+N)}{r_+^2} ,
\end{equation}
where again $\sig=\pm 1$.  This is manifestly non-negative.  Hence the BF bound is respected so we do not predict any instability. This is in agreement with Ref.~\cite{Kunduri:2006}, which proved that $m=0$ tensor perturbations are stable in the full black hole geometry.

\subsubsection{Vector modes}

In contrast with the asymptotically flat case, we are unable to give a simple explicit form for the eigenvalues of $\Ocal{2}$ corresponding to vector modes.  However, we can still prove that for all $N$, for any value of the dimensionless ratio $r_+/l$, the eigenvalues are all non-negative, and hence the effective $AdS_2$ BF bound is respected.  Hence, we do not predict any instability in this sector.

\subsubsection{Scalar modes}\label{sec:gravscalarevalsads}

The analysis proceeds in the same way as in the asymptotically flat case. 

For $\kap=0$, there is a single eigenvalue
\begin{equation}
  \la = L^2\left( \frac{4}{E^2} + 4(N+1)\frac{B^2}{r_+^4}\right)
\end{equation}
which is manifestly positive, and hence there is no instability.

For $\kap=1$, the eigenvalues $\la$ correspond to the eigenvalues of a $4\times 4$ matrix ($3 \times 3$ for $N=1$), and these cannot be found explicitly in a convenient way.  However, plotting these eigenvalues against the dimensionless parameter $r_+/l$ shows immediately that all these eigenvalues lie above the BF bound, and hence there is no instability in this sector.

For $\kappa=2,3,4,\ldots$, the problem reduces to finding eigenvalues of a $6 \times 6$ matrix parameterized by $r_+/l$ ($5\times 5$ for $N=1$).  For each $\kap=2,3,4,\ldots$, there are six real eigenvalues of $\Ocal{2}$.

\begin{figure}
\begin{center}
\begin{minipage}[c]{\textwidth}
  \includegraphics[width=0.49\textwidth]{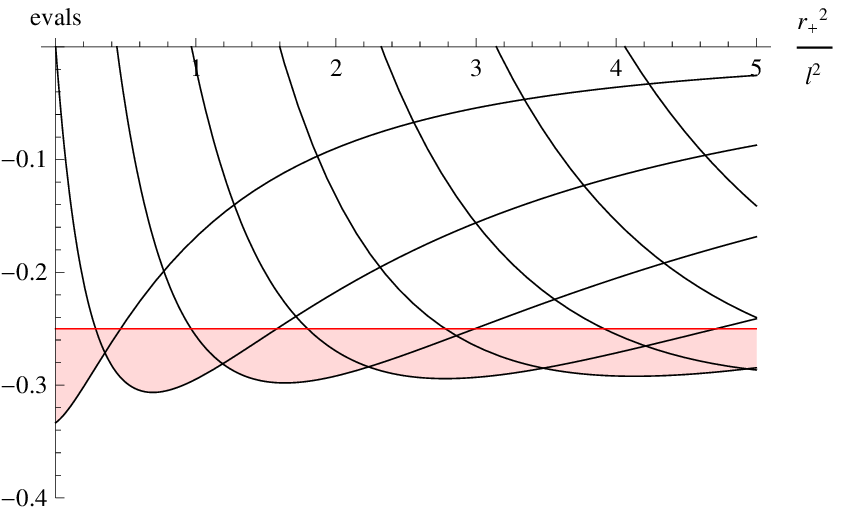}
  \includegraphics[width=0.49\textwidth]{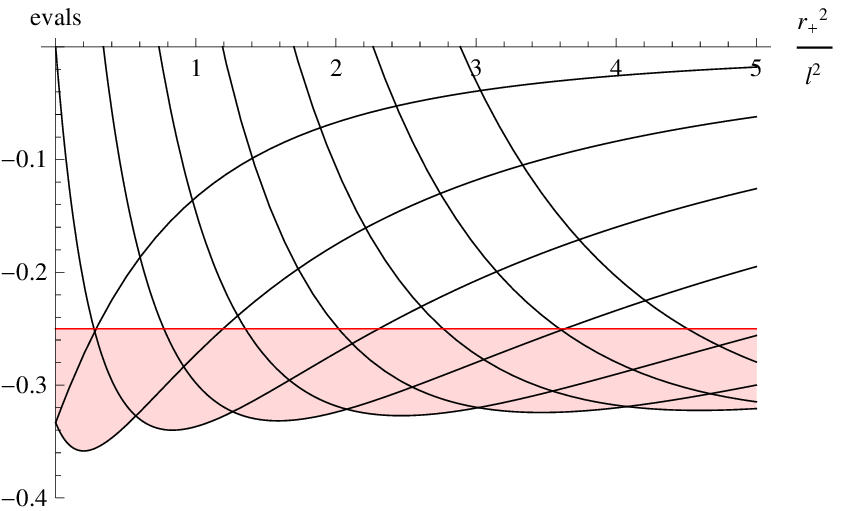}
  \caption{\it\small\label{fig:evalsN=23}Lowest eigenvalues of $\Ocal{2}$ plotted against the size of the $AdS$ black hole ($r_+^2/l^2$), in dimensions $d=7$ (left) and $d=9$ (right). The shaded region corresponds to violation of the effective BF bound. The separate curves shown correspond to $\kap=2,3,4,5,6$, moving from left to right as $\kap$ is increased (the curves that are negative for $r_+/l \rightarrow 0$ are $\kap=2$ on the left and $\kap=2,3$ on the right).  In both cases, there is some mode that violates the BF bound for any black hole size.} 
\end{minipage}
\end{center}
\end{figure}

Our results are easiest to understand for $d \ge 7$ ($N \ge 2$).  Consider first the case $N=2$.  The lowest eigenvalue for each value of $\kap$ is plotted in Figure  \ref{fig:evalsN=23}. We find that there is a violation of the effective BF bound by the lowest $\kap=2$ eigenvalue for sufficiently small $r_+/l$. This makes sense: the eigenvalues here are continuously connected to the eigenvalues in the asymptotically flat case as $r_+/l \rightarrow 0$, and we saw that there is an instability with $\kap=2$ in the asymptotically flat case. Modes with higher $\kap$ are unstable for ranges of $r_+/l$ corresponding to larger black holes.  The ranges for successive values of $\kap$ overlap, and in fact for any $r_+/l$, there exists some $\kap$ corresponding to an unstable mode.  For $N=3$ ($d=9$) the results are similar and are also shown in \ref{fig:evalsN=23}.

We can perform similar studies for higher dimensions $d=11,13,\ldots$, and find results that are qualitatively similar to those for $d=7,9$ (although note that modes with small $\kap$ become stable for small $AdS$ black holes in larger dimensions, however instabilities for higher $\kap$ ensure that such black holes remain unstable).
Therefore our conjecture predicts that all extreme, cohomogeneity-1 MP-AdS black holes with $d \ge 7$ should be unstable against scalar-type gravitational perturbations with $m=0$. We emphasize that this is distinct from the previously discovered superradiant instability.
\begin{figure}
\begin{center}
\begin{minipage}[c]{\textwidth}
  \includegraphics[width=0.49\textwidth]{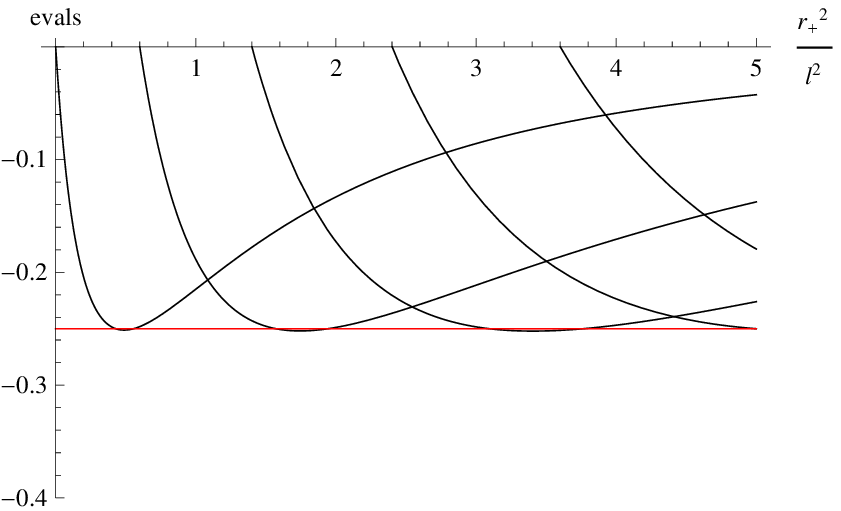}
  \includegraphics[width=0.49\textwidth]{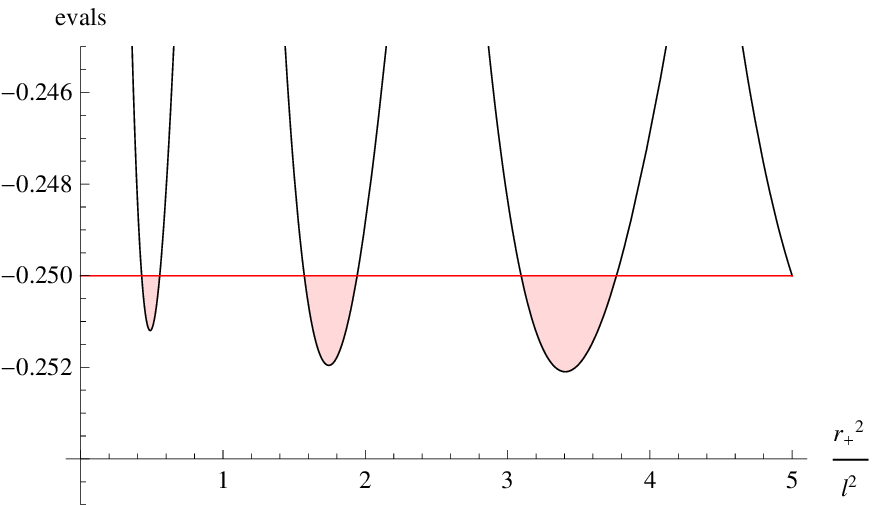}
  \caption{\it\small\label{fig:evalsN=1}Eigenvalues of $\Ocal{2}$ plotted against the size of the $AdS$ black hole ($r_+^2/l^2$), in dimension $d=5$ (the right hand graph is a zoomed version of the left hand one).  The separate curves shown correspond to $\kap=2,3,4,5,6$, moving from left to right as $\kap$ is increased.  We find that the generalized BF bound $\la \geq-1/4$, shown by the horizontal line, is violated by a small amount in various small ranges of black hole size.  As $\kap$ is increased, the violation of the BF bound occurs for increasingly large black holes.} 
\end{minipage}
\end{center}
\end{figure}

For $d=5$ ($N=1$), we plot the lowest eigenvalue of $\Ocal{2}$ with given $\kappa$ in Figure \ref{fig:evalsN=1}.
For $\kap=2$, there is a violation of the effective BF bound for $0.43<r_+^2/l^2<0.56$.  The violation is small: by less than $1\%$.  Modes with higher $\kap$ are also unstable in particular small ranges of the black hole size, but for increasingly large black holes as $\kap$ increases.  We do not have a good explanation for why these unstable modes are found only in these small ranges. The fact that the bound is violated only by a small amount may imply that the instability appears much closer to extremality than anything in Table \ref{tab:jorge} so confirming our conjecture in this case may require a numerical study of the full, \emph{extremal} black hole solution.  As we can only give a sufficient condition for instability, not a necessary one, it might be the case that the full extremal black hole solution is unstable against $m=0$ perturbations for any $r_+/l$ above a certain lower bound (we know that an instability is not present for the asymptotically flat case $r_+/l \rightarrow 0$). 

\subsection{Electromagnetic Perturbations}

\subsubsection{Introduction}

Recall that an instability of the near-horizon geometry under electromagnetic perturbations  corresponds to an eigenvalue of $\Ocal{1}$ being less than $-1/4$, and the eigenvectors of $\Ocal{1}$ are vectors $Y_\mu$ on $S^{2N+1}$. Just as in the gravitational case, we can decompose these into parts parallel and perpendicular to $\CP{N}$ and then decompose these parts into scalar and vector harmonics on $\CP{N}$.  As things are simpler here, we can consider both asymptotically flat and asymptotically $AdS$ black holes together (the asymptotically flat case corresponds to setting $\tfrac{r_+^2}{l^2} = 0$ below).  We find no evidence of any instability in either of these cases. Once again, we restrict attention to modes with $m=0$ since these are the ones relevant to our conjecture.

\subsubsection{Vector modes}\label{sec:emvectors}
For eigenvectors $Y_\mu$ built from vector harmonics on $\CP{N}$ (which exist only for $N>1$), we find eigenvalues
\begin{equation}
 \la = \frac{2\big(\kap^2+(N+3)\kap + 2(N+1)\big)}{(N+1)\left(N+(N+2)\tfrac{r_+^2}{l^2}\right)},
\end{equation}
where $\kap$ is a non-negative integer.  These are all positive so there is no instability.

\subsubsection{Scalar modes}\label{sec:emscalars}

For $Y_\mu$ built from scalar harmonics on $\CP{N}$ (labelled by a non-negative integer $\kap$), there are two cases to consider separately.  For $\kap=0$, there is a positive single eigenvalue:
\begin{equation}\label{eqn:emevals0}
  \la = \frac{2N\left(1+\tfrac{r_+^2}{l^2}\right)}{N+(N+2)\tfrac{r_+^2}{l^2}}
\end{equation}
For $\kap\geq 1$, there are three eigenvalues for each $\kap$, given by
\begin{equation}
  \la = \frac{2\kap(\kap+N)}{(N+1)\left(N+(N+2)\tfrac{r_+^2}{l^2}\right)},
\end{equation}
and
\begin{equation}
  \la = \frac{\tfrac{2\kap(\kap+N)}{N+1}+N + (N+1)\tfrac{r_+^2}{l^2}
              \pm \sqrt{4\kap(\kap+N)\left(1 + \left(\tfrac{N+2}{N+1}\right) \tfrac{r_+^2}{l^2}
                                     \right)
                        + \left(N+(N+1)\tfrac{r_+^2}{l^2}\right)^2
                       }
             }
             {N+(N+2)\tfrac{r_+^2}{l^2}}.
\end{equation}
Two of these are positive, but the third can sometimes be negative.  In order to check whether the effective $AdS_2$ BF bound $\la \geq -1/4$ is violated, we plotted this eigenvalue against the $AdS_d$ black hole size $r_+/l$, finding that there is no violation of the effective BF bound for any $N$ or $\kap$.

In the asymptotically flat case, the eigenvalues are again very simple, reducing to $\lambda=2$ when $\kappa=0$ and for $\kappa \ge 1$ they are
\begin{equation} \label{eqn:asflatmaxevals}
  \frac{2\kap(\kap+N)}{N(N+1)},\quad
  \frac{2(\kap+N)(\kap+N+1)}{N(N+1)}, \quad
  \frac{2\kap(\kap-1)}{N(N+1)}.
\end{equation}

\subsection{Dual operators and conformal dimensions}\label{sec:kerrcft}

It has been conjectured that there exists a chiral CFT dual to the NHEK geometry \cite{kerrcft} and that a non-chiral CFT governs excitations away from extremality \cite{Bredberg:2009pv}. States of the latter fill out representations of two copies of the Virasoro algebra. Assuming that CFT operator dimensions are related to the decay rate of fields in $AdS_2$ in the usual way, then equation (\ref{eqn:modefreq}) gives the conformal weights with respect to one of these algebras.
In general, these turn out to be complex, which may be a problem for the Kerr-CFT conjecture. However, the results of Refs \cite{Amsel:2009ev,Dias:2009ex} imply that operators dual to {\it axisymmetric} gravitational perturbations are particularly simple, with integer conformal weights $\Delta_+ = l+1 $ where $l=2,3,\ldots$ labels the harmonic on $\Hcal = S^2$. 

It has been suggested that the Kerr-CFT conjecture can be extended to the Myers-Perry black holes \cite{popekerrcft} so it is interesting to use our results to compute operator weights for this case too. Consider a cohomogeneity-1 extreme Myers-Perry black hole. The operator $\Ocal{2}$ governing gravitational perturbations of the near-horizon geometry appears very complicated.  It is striking that its eigenvalues are all rational numbers (for asymptotically flat black holes\footnote{
In the asymptotically AdS case, the eigenvalues generically are all irrational but this case seems less interesting for the present discussion since there always is a superradiant instability \cite{Kunduri:2006}.}). 

For $d>5$ we have seen that our conjecture predicts an instability so presumably a CFT dual does not exist (or is also unstable). So consider the case $d=5$ ($N=1$). In this case, only scalar-type gravitational perturbations exist. Again, if $m \ne 0$ then there are complex conformal weights so consider the modes with $m=0$ covered by our conjecture. For $\kappa=0$ we have $\Delta_+ = 3$. The $\kappa=1$ harmonics give
\be
 \Delta_+ = 2, 3, 4
\ee
For $\kap=2,3,4, \ldots$, we find
\be
 \Delta_+ = \kappa-1,\, \kappa, \, \kappa+1, \, \kappa+2,\, \kappa+3
\ee
Hence if there is a CFT description that obeys the usual AdS/CFT rules then the $m=0$ gravitational perturbations give rise to five infinite families of operators with {\it integer} conformal weights, just as for NHEK.  This suggests that some symmetry is protecting the weights of operators dual to $m=0$ gravitational perturbations.

A massless scalar field would give operators with $\Delta_+ = \kap+1$ for $\kap=0,1,2\ldots$, while equations (\ref{eqn:emevals0},\ref{eqn:asflatmaxevals}) imply that Maxwell perturbations correspond to operators with weights
\begin{equation}
  \Delta_+ = 2 \qquad (\kap = 0) \eqand \Delta_+ = \kappa, \, \kappa+1, \, \kappa+2 \qquad (\kap \geq 1).
\end{equation}
For $N>1$, if we ignore the instability and calculate $\Delta_+$ formally for gravitational perturbations (for stable modes) then the results are generically irrational.  

Following the appearance of the first version of the present paper, Ref. \cite{Murata:2011my} found that all of these results for conformal weights are unchanged if one considers the more general case of the near-horizon geometry of a 5d extreme Myers-Perry black hole with {\it unequal} angular momenta.
                                   
\section{Instabilities from near-horizon geometries}\label{sec:nhinstab}

\subsection{Introduction}

Does an instability of the near-horizon geometry imply the existence of an instability of the full spacetime?  We conjectured in the introduction that this was the case for a particular class of perturbation modes and explained how extreme Kerr is consistent with the conjecture. In Section \ref{sec:nhmp} we have shown that our conjecture predicts an instability for certain Myers-Perry black holes, and this prediction is confirmed by studies of perturbations of the full black hole geometry. 

In this section, we will present some ideas that explain why our conjecture appears to work. In the case of a scalar field, we shall sketch a proof of the conjecture. We shall present some evidence suggesting that the method of proof in the scalar field case might also generalize to gravitational perturbations. 

\subsection{Scalar field instabilities}

Consider an uncharged, scalar field $\Psi$ of mass $M$ in the extreme planar Reissner-Nordstrom-AdS black hole background in arbitrary dimension $d\geq 4$.  This has a near-horizon geometry of the form $AdS_2 \times \Rbb^{d-2}$.  So, in the language described above, we have $\Hcal=\Rbb^{d-2}$ here.

As before, we can Fourier analyze on $\Rbb^{d-2}$ to reduce the scalar field equation of motion to that of a massive scalar in $AdS_2$.  The BF bound (\ref{eqn:ads2stabbound}) associated to the $AdS_2$ is more restrictive than that associated to the asymptotic $AdS_d$ geometry. Numerical work \cite{Hartnoll:2008kx,Denef:2009tp} suggests that if the scalar field violates the $AdS_2$ BF bound then the scalar field is unstable in the full black hole geometry (even when the $AdS_d$ BF bound is respected). Moreover, it has been proved \cite{Dias:2010ma} that if the $AdS_2$ BF bound is satisfied then the scalar field is stable in the full black hole geometry, i.e., stability of the near-horizon geometry implies stability of the full black hole. Here we will prove that {\it instability} of the near-horizon geometry implies instability of the full black hole, in agreement with our conjecture.

Consider an extreme static black hole with geometry
\begin{equation}\label{eqn:staticgeom}
  ds^2 = -f(r) dt^2 + f(r)^{-1} dr^2 + r^2 d\Sigma_k^2.
\end{equation}
where $d\Sigma_k^2$ is the metric on a unit sphere if $k=1$, a unit hyperboloid if $k=-1$ and flat if $k=0$.  This metric encompasses the Schwarzschild(-AdS) and Reissner-Nordstrom(-AdS) black holes with various horizon topologies.

As the black hole is extreme, we can assume that it has a degenerate horizon at $r=r_+$, and hence that
\begin{equation}
  f(r) = \frac{(r-r_+)^2}{L^2} + \mathcal{O} (r-r_+)^3.
\end{equation}
The near-horizon geometry is then $AdS_2 \times \Sigma_k$ where the $AdS_2$ has radius $L$.

In the full spacetime, the equation of motion of a scalar field $\Psi$ of mass $M$ can be written
\begin{equation} \label{Aeq}
 -\frac{\partial^2 \Psi}{\partial t^2} = {\cal B} \Psi,
\end{equation}
where
\begin{equation}
 {\cal B}\Psi \equiv f \left[ - \frac{1}{r^{d-2}} \pd_r \left( r^{d-2} f \pd_r \Psi \right) 
                        - \frac{1}{r^2} \hat{\nabla}^2 \Psi + M^2 \Psi \right],
\end{equation}
with $\hat{\nabla}$ the connection on $\Sigma_k$.  Now define the following inner product between functions defined on a surface of constant $t$ outside the horizon:
\begin{equation}
 ( \Psi_1, \Psi_2 ) = \int_{r_+}^\infty dr \, d\Sigma_k \, r^{d-2} f^{-1} \Psi_1 \Psi_2 .
\end{equation}
We impose boundary conditions that the functions of interest must decay sufficiently fast for this integral to converge at $r=\infty$, and they must vanish at least as fast as $(r-r_+)$ as $r \rightarrow r_+$ in order that the integral converges at $r=r_+$.  Now, if our functions decay fast enough at infinity, then ${\cal B}$ is self-adjoint with respect to this inner product.\footnote{Note that this is different to the self-adjointness of operators discussed in Section \ref{sec:decomposition}; we are integrating over the exterior region of the full spacetime, not just over the manifold $\Hcal$.}

We can estimate the lowest eigenvalue $\lambda_0$ of $\Bcal$ using the Rayleigh-Ritz method, noting that
\begin{equation} \label{rayleighritz}
 \lambda_0 \le \frac{ (\Psi, {\cal B} \Psi )}{(\Psi,\Psi)},
\end{equation}
for any function $\Psi$ satisfying the boundary conditions.

Suppose that $\lambda_0$ is negative, with $\Psi_0$ the associated eigenfunction. Then (\ref{Aeq}) has solutions 
\begin{equation}
  \Psi(t,r,x) = e^{\pm \sqrt{-\lambda_0 } t} \Psi_0.
\end{equation}
From the form of ${\cal B}$, it is easy to show that near $r=r_+$, the eigenfunction behaves as \begin{equation}
  \Psi_0 \sim \exp \left(-\frac{\sqrt{-\lambda_0} L^2}{r-r_+}\right) .
\end{equation}
Transforming to ingoing Eddington-Finkelstein coordinates ($dv = dt + dr/f$ so $t \sim v + L^2/(r-r_+)$ near $r=r_+$) reveals that the solution $e^{+ \sqrt{-\lambda_0 } t} \Psi_0$ is regular at the future horizon.  This grows exponentially with time, and hence represents an instability of the scalar field in the black hole background.

The idea now is to show that violation of the $AdS_2$ BF bound (\ref{eqn:ads2stabbound}) implies the existence of a trial function $\Psi$ with $(\Psi, {\cal B} \Psi) < 0$. This implies that $\lambda_0$ must be negative, hence the scalar field is unstable and the conjecture is proved.

To see how this works, consider the case of a 4d extreme Reissner-Nordstrom-AdS black hole, for which
\begin{equation}
 f(r) = \left( 1-\frac{r_+}{r} \right)^2 \left(k + \frac{3 r_+^2  + 2 r r_+ + r^2}{l^2} \right), 
\end{equation}
where $\ell$ is the $AdS_4$ radius.  This has a near-horizon geometry with
\begin{equation}
 \frac{1}{L^2} = \frac{6}{l^2} + \frac{k}{r_+^2}.
\end{equation}
Consider the following trial function (motivated by a similar example in Ref.\ \cite{Dias:2010ma}) 
\begin{equation} \label{trial}
 \Psi(r) = \frac{(r-r_+) l^{9/2}}{r^4 (r-r_+ + \epsilon l)^{3/2}},
\end{equation}
with $\epsilon >0$. This satisfies the boundary conditions required for self-adjointness of ${\cal B}$. As $\epsilon \rightarrow 0$, this gives
\begin{equation}
 (\Psi, {\cal B} \Psi ) 
    = \int_{r_+}^\infty dr \, d\Sigma_k \, r^{2} \left( f (\partial_r \Psi)^2 + M^2 \Psi^2 \right) 
    = V_k \left( M^2 + \frac{1}{4 L^2} \right) \frac{\ell^9}{r_+^6} \log \left( \epsilon^{-1} \right)
      + {\ldots}
\end{equation}
where the ellipsis denotes terms subleading in $\epsilon$, and $V_k$ is the volume of $\Sigma_k$.  The $AdS_2$ BF bound states that the quantity in brackets on the RHS should be non-negative.\footnote{
More precisely: this is the BF bound for modes which are homogeneous on $\Sigma_k$.}
From the above expression we see that ${\cal B}$ admits a negative eigenvalue if this bound is violated.  Hence there is an instability of the scalar field when the $AdS_2$ BF bound is violated. 
The argument generalizes easily to $d>4$.\footnote{
Ref.  \cite{Dias:2010ma} proved that {\it stability} of the near-horizon geometry implies stability of the full black hole for $k=-1,0$. Combining this with our result, we learn that, for these cases, the full black hole is stable if, and only if, its near-horizon geometry is stable. For $k=1$, stability of the near-horizon geometry is not sufficient to guarantee stability of the full black hole because the $AdS_2$ BF bound can be less restrictive than the $AdS_d$ bound.}

A similar example is the cohomogeneity-1 Myers-Perry-AdS black hole discussed in section \ref{sec:nhmp}. For large black holes, the $AdS_2$ BF bound is more restrictive than that of $AdS_d$. 
Hence a scalar field can violate the $AdS_2$ BF bound but respect the $AdS_d$ BF bound.  Ref.\ \cite{Dias:2010ma} studied the case of a scalar field invariant under $\partial/\partial \psi$ (i.e.\ those modes corresponding to $m=0$ in Section \ref{sec:scalarfield}) and presented numerical evidence that such a scalar field is indeed unstable if its mass lies between the two BF bounds. Furthermore, it was proved that the scalar field (with $m=0$) is stable if it respects both bounds.

This example also can be understood using the argument above.  Even though the black hole is rotating, the fact that the scalar field is invariant under $\partial/\partial \psi$ implies that its equation of motion takes the form (\ref{Aeq}). The only difference is the form of ${\cal B}$:
\begin{equation}
 {\cal B}\Psi = \frac{V(r)}{h(r)^2}  
   \left[ - \frac{1}{r^{d-2}} \pard{r} \left( r^{d-2} V(r) \frac{\pd \Psi}{\pd r} \right) 
          - \frac{1}{r^2} \hat{\nabla}^2 \Psi + M^2 \Psi \right],
\end{equation}
where $\hat{\nabla}$ is the connection of the metric on $\CP{N}$. $\Bcal$ is self-adjoint with respect to the scalar product
\begin{equation}
 (\Psi_1,\Psi_2) = 2 \pi \int_{r_+}^\infty dr \, d\hat{\Sigma}_N\, r^{d-2} \frac{h(r)^2}{V(r)} \Psi_1\Psi_2,
\end{equation}
where $d\hat{\Sigma}_N$ is the volume element on $\CP{N}$.  Consider, for simplicity, the five-dimensional case (where $N=1$).  We can use the trial function (\ref{trial}) with the modification $r^4 \mapsto r^5$ (to improve the convergence at $r=\infty$).  The result is the same: $(\Psi, {\cal B} \Psi)$ is proportional to $\log (\epsilon^{-1})$ with a coefficient of proportionality that is negative if, and only if, the $AdS_2$ BF bound is violated.  Hence we have proved that the scalar field is unstable in the extreme black hole geometry if it violates the $AdS_2$ BF bound, in agreement with our conjecture.

Now recall from the introduction that for the extreme Kerr black hole, we know that instability of the near-horizon geometry does not always imply instability of the full black hole. Even for a scalar field, there exist modes that violate the effective BF bound in the near-horizon geometry \cite{Bardeen:1999px}. So how does the above argument fail for Kerr? The key step above was to impose a symmetry condition on the scalar field that makes its equation of motion take the form  (\ref{Aeq}), in which first time derivatives are absent. For Kerr, eliminating first time derivatives implies that the scalar field must be axisymmetric, and axisymmetric modes do respect our conjecture. 

More generally, if we consider an extreme black hole with metric (\ref{ADM}) then the necessary and sufficient condition for the equation of motion of a massive scalar field to reduce to (\ref{Aeq}) is 
\begin{equation}
 N^I(x) \frac{\partial}{\partial \phi^I} \Psi = 0.
\end{equation} 
If we Fourier analyze $\Psi \propto e^{i m_I \phi^I}$ for integers $m_I$ then this equation reduces to the axisymmetry condition (\ref{eqn:axicondition}) in the conjecture that we made in the introduction. If this condition is satisfied then the argument we have sketched above should apply. This explains why our conjecture should work for scalar fields.

\subsection{Gravitational perturbations}

We have sketched an argument that explains why a scalar field instability in the near-horizon geometry of an extreme black hole implies an instability of the full black hole, provided the scalar field satisfies the symmetry condition (\ref{eqn:axicondition}).  We are really interested in linearized gravitational perturbations.  If we attempt to repeat the same argument, we would need to convert the equations governing gravitational perturbations to something of the form 
\begin{equation} \label{Aeq2}
 -\frac{\partial^2 \Psi_\alpha}{\partial t^2} = {\cal A}_\alpha{}^\beta \Psi_\beta
\end{equation}
with $\Psi_\alpha$ a vector encoding the perturbation, and $ {\cal A}_\alpha{}^\beta$ a matrix operator self-adjoint with respect to a suitable inner product.  Can this be done?  For axisymmetric (i.e.\ $m=0$) metric perturbations of the Kerr black hole, in a certain gauge, it can indeed: a variational formula analogous to (\ref{rayleighritz}) is given in Ref. \cite[\S 114]{chandra}.  Hence the extreme Kerr black hole should obey our conjecture and, as we discussed in the introduction, it does.

What about higher dimensions? Can we bring the equations governing gravitational perturbations of, for example, a Myers-Perry black hole to the form (\ref{Aeq2}), provided the perturbation satisfies the symmetry condition (\ref{eqn:axicondition})? Evidence that this is indeed possible comes from recent work \cite{Dias:2010eu} on instabilities of cohomogeneity-1 MP black holes.  This work considered metric perturbations satisfying (\ref{eqn:axicondition}).  In the cases for which an instability was found, the time dependence was $e^{-i \omega t}$ where $\omega$ has positive imaginary part. In general, one would expect $\omega$ to be complex but it turned out that unstable modes had purely imaginary $\omega$. This would be explained if perturbations were governed by an equation of the form (\ref{Aeq2}) (with $ {\cal A}_\alpha{}^\beta$ self-adjoint), which predicts that $\omega^2$ should be real.

Perturbations of Myers-Perry black holes with a single non-vanishing angular momentum have also been considered \cite{Dias:2009iu}. Again, perturbations satisfying (\ref{eqn:axicondition}) were considered ((\ref{eqn:axicondition}) reduces to $m_1=0$ in this case). The critical mode associated to the onset of instability was identified. This mode has zero frequency, which suggests that unstable modes should have purely imaginary frequency (if unstable modes had $\omega$ with a non-zero real part then there is no reason why the mode at the threshold of instability should have $\omega=0$ rather than $\omega$ some non-zero real number).  Again, this suggests that a formula of the form (\ref{Aeq2}) exists for this situation.

In these two examples, it appears that the condition (\ref{eqn:axicondition}) is indeed sufficient to obtain an equation of the form (\ref{Aeq2}) governing gravitational perturbations (in a certain gauge). This is encouraging evidence that it will indeed be possible to demonstrate that an instability of the near-horizon geometry of an extreme black hole will imply instability of the full black hole provided this symmetry condition is respected. Future work will investigate these issues in more detail.

\section*{Acknowledgments}

We are very grateful to Jorge Santos for supplying us with the numerical results quoted in Table \ref{tab:jorge} and to Keiju Murata for discussions and pointing out several errors in a previous version of this paper.
We are also  grateful to Oscar Dias and Pau Figueras for discussions and to Mahdi Godazgar for comments on a draft of this paper.  MND is supported by STFC, HSR is a Royal Society University Research Fellow.

\appendix
\section{Computations for general near-horizon geometries} \label{app:nhframe}
In this appendix, we explain the calculations required to obtain the results presented in Section \ref{sec:decoupling} for a general metric ansatz (\ref{nhgeom}) including all known near-horizon geometries.

\subsection{Metric and null frame}
Consider a near-horizon geometry of the form (\ref{nhgeom}), with $n$ rotational Killing vectors $\pd/\pd \phi^I$, and indices $I,J,\ldots=2,\ldots n+1$ and $A,B,\ldots = n+2,\ldots d-1$.  We think of this as a fibration over $AdS_2$ of some manifold $\Hcal$ with metric
\begin{equation}
  d\hat{s}^2 = g_{IJ}(y)d\phi^I d\phi^J + g_{AB}(y) dy^A dy^B 
             = \hat{g}_{\mu\nu}d\hat{x}^\mu d\hat{x}^\nu .
\end{equation}
The rotation of the black hole is described by the constants $k^I$.  It is useful to define a (Killing) vector field $k = k^I \pard{\phi^I}$. 

In Ref.\ \cite{decoupling} we derived decoupled equations for gravitational perturbations and Maxwell test fields in the background of Kundt spacetimes, using the higher-dimensional Geroch-Held-Penrose (GHP) formalism \cite{higherghp}.  In this section, we show that all metrics of the form (\ref{nhgeom}) are (doubly) Kundt spacetimes, and compute the relevant equations in these particular cases.  To do this, we will make use of the notation and results of the higher-dimensional GHP formalism \cite{higherghp}, and readers of this appendix may wish to first familiarise themselves with the basic definitions made therein.  The results obtained, for use in the rest of the paper, will be expressed in notation independent of this formalism.

We work in a null frame
\begin{align}
  \lb  &= e_0 = e^1 = \tfrac{1}{\sqrt{2}} L(y) \left(-R dT + \tfrac{dR}{R} \right),\nn\\
  \nb  &= e_1 = e^0 =  \tfrac{1}{\sqrt{2}} L(y) \left(R dT + \tfrac{dR}{R} \right) ,\nn\\
  m_\Ih  &= e_\Ih = e^\Ih = \hat{e}_{\Ih I}\left( d\phi^I - k^I R dT \right),\nn\\
  m_\Ah &= e_\Ah = e^\Ah =\hat{e}_\Ah,
\end{align}
where $\hat{e}$ are vielbeins for $\Hcal$.  Indices $\Ih,\Jh,\dots=2,\dots n+1$ are frame indices in the Killing directions, while $\Ah,\Bh,\dots = n+2,\dots d-1$ are frame indices in the non-Killing directions.

With indices raised this gives
\begin{align}
  e_0   &= \frac{1}{L\sqrt{2}} \left( \frac{1}{R}\pard{T}+k^I\pard{\phi^I}+R\pard{R} \right),\nn\\
  e_1   &= \frac{1}{L\sqrt{2}} \left(-\frac{1}{R}\pard{T}-k^I\pard{\phi^I}+R\pard{R} \right),\nn\\
  e_\Ih   &= \hat{e}_\Ih^I \pard{\phi^I }, \nn\\
  e_\Ah &= \hat{e}_\Ah.
\end{align}

Using the Cartan equations $de_a + \om_{ab}\wedge e^b = 0$ we find that the spin connection is given by
\begin{align}
  \om_{01} &= \tfrac{1}{L\sqrt{2}} (e_0 - e_1) - \tfrac{1}{2L^2}(k^I\hat{e}_{I\Ih}) e_\Ih , &
  \om_{0\Ih} &= - \tfrac{1}{2L^2}(k^I\hat{e}_{I\Ih}) e_0, \nn\\
  \om_{0\Ah} &= \tfrac{1}{L}(dL)_\Ah \, e_0 , &
  \om_{1\Ih} &= + \tfrac{1}{2L^2}(k^I\hat{e}_{I\Ih}) e_1,\nn\\
  \om_{1\Ah} &= \tfrac{1}{L}(dL)_\Ah \, e_1 , &
  \om_{\Ih\Jh} &= -\hat{e}_\Jh.\left[(e_\Ah.\nabla)\hat{e}_\Ih\right] e_\Ah,\nn\\
  \om_{\Ah\Bh} &= \hat{\om}_{\Ah\Bh}, &
  \om_{\Ah \Ih} &= 0.
\end{align}
This is sufficient to give us the GHP optical scalars for the spacetime, which in the notation of \cite{higherghp} read
\begin{equation}
  \kap_i = \kap'_i = 0, \qquad
  \rho_{ij} = \rho'_{ij} = 0, \qquad
  \tau_i = \frac{ k_i - d(L^2)_i}{2L^2}
\end{equation}
where $i,j\dots = 2,\dots,d-1$ are frame indices on the $d-2$ spacelike dimensions (or equivalently on $\Hcal$).  This implies that both $\lb$ and $\nb$ define geodesic, non-expanding, non-shearing, non-twisting null congruences, and hence that this is a (doubly) Kundt spacetime.  All Kundt spacetimes are algebraically special  \cite{Ricci}, and by a simple extension of this argument it can be seen that all doubly Kundt spacetimes are of algebraic Type D, in the sense defined by Ref.\ \cite{cmpp} (and reviewed in \cite{higherghp}).

For this metric, the GHP derivative operators \cite[\S 2.3]{higherghp}, acting on a GHP scalar $T_{i_1\dots i_s}$ of boost weight $b$ and spin $s$, are
\begin{align}
  \tho T_{i_1\dots i_s}
       &= \frac{1}{L\sqrt{2}} \left( \frac{1}{R}\pard{T}+k.\pard{\phi}
                                    + R\pard{R} - b\right)T_{i_1\dots i_s},\label{eqn:tho}\\
  \tho' T_{i_1\dots i_s}
       &= \frac{1}{L\sqrt{2}} \left(-\frac{1}{R}\pard{T}-k.\pard{\phi}
                                    + R\pard{R} + b\right)T_{i_1\dots i_s},\label{eqn:thop}\\
  \eth_j T_{i_1\dots i_s} 
       &= \left(\nablah_j - \frac{b}{2L^2} k_j \right) T_{i_1\dots i_s}\label{eqn:eth}
\end{align}
where $\nablah$ is the covariant derivative on $\Hcal$.  In Ref.~\cite{decoupling} it was shown that test scalar fields $\phi$, Maxwell fields $\vphi_i = F_{0i}$ and gravitational perturbations $\Om_{ij} = C_{0i0j}$ of Kundt spacetimes such as this can be described by the equations
\begin{equation} \label{eqn:scalarperts}
  (2\tho'\tho + \eth_i\eth_i + \rho'\tho  -2\tau_i\eth_i - M^2)\phi = 0,
\end{equation}
\begin{equation}\label{eqn:maxperts}
    \left(2\tho'\tho + \eth_j\eth_j + \rho'\tho -4\tau_j\eth_j 
                                   + \Phi-\tfrac{2d-3}{d-1}\La\right)\vphi_i 
             + (- 2\tau_i\eth_j + 2\tau_j \eth_i + 2\Phis_{ij} + 4\Phia_{ij})\vphi_j = 0,
\end{equation}
and
\begin{multline}\label{eqn:gravperts}
  \left(2\tho'\tho+ \eth_k \eth_k + \rho'\tho - 6\tau_k\eth_k  + 4\Phi 
        - \tfrac{2d}{d-1} \La \right) \Om_{ij}\\
  + 4\left(\tau_k\eth_{(i|}- \tau_{(i|}\eth_k + \Phis_{(i|k} + 4\Phia_{(i|k}\right) \Om_{k|j)} 
  + 2\Phi_{ikjl} \Om_{kl} = 0.
\end{multline}

\subsection{General fields}
Now consider a GHP covariant field $T_{i_1\dots i_s}$ of boost weight $b$ and spin $s$ (in the sense of Definition 1 of \cite{higherghp}).  That is, take $T$ to be one of $\phi$, $\vphi_i$, $\Om_{ij}$, which have $(b,s)=(0,0), (1,1), (2,2)$ respectively.

Consider a separable ansatz
\begin{equation}\label{eqn:separation}
  T_{i_1\dots i_s}(T,R,\phi^I,y^A) = \chi_b(T,R) \, Y_{i_1\dots i_s}(\phi^I,y^A),
\end{equation}
where $\chi_b$ has boost weight $b$, and $Y$ has boost weight 0.  We think of $\chi_b$ as a field on $AdS_2$, and $Y$ as a tensor on $\Hcal$.  Eventually it will be useful to move away from the null frame, so let $\mu,\nu,\dots$ be coordinate indices on $\Hcal$.

Note that the GHP derivative $\eth_i$ reduces to the standard covariant derivative on $\Hcal$ when acting on boost weight zero fields such as $Y$.  Hence, given a decomposition of the form (\ref{eqn:separation}), we see that equation (\ref{eqn:eth}) reduces to
\begin{equation}
  \eth_j T_{i_1\dots i_s} = \chi_b \nablah_j Y_{i_1\dots i_s}
                            - Y_{i_1\dots i_s} \frac{b}{2L^2} k_j \chi_b.
\end{equation}

We can take Fourier expansions of the dependence of $Y$ on the coordinates $\phi^I$, of the form $Y \sim e^{i m_I \phi^I}$, which is equivalent to the statement that the Lie derivative of $Y$ with respect to $\pd/\pd \phi^I$ is given by
\begin{equation}
  (\Lcal_I Y)_{\mu_1\dots \mu_s} = i m_I Y_{\mu_1\dots \mu_s},
\end{equation}
and hence
\begin{equation}
  (\Lcal_k Y)_{\mu_1\dots \mu_s} = i k.m Y_{\mu_1\dots \mu_s},
\end{equation}
where $k.m \equiv k^I m_I$.  For the three different kinds of field, this implies that
\begin{align}
  k.\nablah Y &= i k.m Y, \\
  k.\nablah Y_\mu &= ik.m Y_\mu - (\nablah_\mu k^\nu) Y_\nu\\
  k.\nablah Y_{\mu\nu} &= ik.m Y_{\mu\nu} - 2(\nablah_{(\mu|} k^\rho) Y_{\rho|\nu)}.
\end{align}

Recall now the equation of motion $(D^2-\mu^2)\chi_b = 0$ for a charged massive scalar field $\chi$ on a unit radius $AdS_2$ space, described by the metric (\ref{ads2}), where the charged covariant derivative $D$ was defined by \eqref{eqn:ads2deriv}.  Explicitly, this equation of motion reads 
\begin{equation}
  - \frac{1}{R^2} \frac{\pd^2 \chi}{\pd T^2} - \frac{2iq}{R} \frac{\pd\chi}{\pd T}
  + \frac{\pd}{\pd R} \left( R^2 \frac{\pd\chi}{\pd R} \right) + (q^2-\mu^2) \chi = 0.
\end{equation}

Using the equations (\ref{eqn:tho},\ref{eqn:thop}), it can then be shown that
\begin{equation}
  2\tho'\tho T_{i_1\dots i_s} = \frac{1}{L^2} \left[ D^2 \chi_b + iq \chi_b \right] Y_{i_1\dots i_s}
\end{equation}
where $D^2$ is the square of the $AdS_2$ operator (\ref{eqn:ads2deriv}) and $q = ib+k.m$.  Also, we have
\begin{equation}
  \eth_j \eth_j T_{i_1\dots i_s} 
             = \frac{\chi_b}{L^2} \left[ \frac{b^2}{4L^2} k.k - b k.\nablah  
                                         + L^2 \nablah^2 \right] Y_{i_1\dots i_s}
\end{equation}
and
\begin{equation}
  -2(b+1) \tau_j \eth_j T_{i_1\dots i_s} 
               = \frac{\chi_b}{L^2}\left[ \frac{b(b+1)}{2L^2} (k.k) - (b+1) k.\nablah 
                                          + (b+1) d(L^2).\nablah \right] Y_{i_1\dots i_s}.
\end{equation}

Now consider the boost weight zero Weyl tensor components $\Phi$, $\Phi^S_{ij}$, $\Phi^A_{ij}$, $\Phi_{ijkl}$ that appear in equations (\ref{eqn:scalarperts}-\ref{eqn:gravperts}).  Recall that the NH geometry is an Einstein spacetime with Ricci tensor $R_{ab} = \La g_{ab}$.  Given this, we can use equation (4.5) of \cite{higherghp} to write
\begin{equation}
  \Phi_{ijkl} = \Rh_{ijkl} - \tfrac{2\La}{d-1} \del_{[i|k}\del_{|j]l}
\end{equation}
where $\Rh_{ijkl}$ is the Riemann tensor of $\Hcal$.  Taking traces of this with the metric on $\Hcal$ implies that
\begin{equation}
  2\Phis_{ij} =  - \Rh_{ij} + \tfrac{d-3}{d-1} \La \del_{ij} ,\qquad
  2\Phi = - \Rh + \tfrac{(d-2)(d-3)}{d-1} \La .
\end{equation}
The remaining components $\Phia_{ij}$ are not related to the curvature of $\Hcal$, but instead can be computed using equation (NP4) of \cite{higherghp}, giving
\begin{equation}
  2\Phia_{ij} = -2\eth_{[i} \tau_{j]} = -(d\tau)_{ij} 
              = -\left( \frac{dk}{2L^2} - \frac{(dL^2)\wedge k}{2L^4} \right)_{ij}.
\end{equation}

\subsection{Separating equations}
In the case of a scalar field, $b=s=0$, and this is enough to allow us to immediately write out equation (\ref{eqn:scalarperts}) as 
\begin{equation}
  Y \left[(D^2 - q^2)\chi_0\right] = \chi_0\left[- \nablah^\mu (L^2 \nablah_\mu Y) 
                                           - (k.m)^2 Y + M^2 L^2 Y \right]
\end{equation}
and hence we can separate variables to obtain
\begin{equation}
    (D^2 - q^2 - \la)\chi_0(T,R) =0
\end{equation}
and
\begin{equation}
 \left[ - \nablah^\mu (L(y)^2 \nablah_\mu ) - (k.m)^2 + M^2 L(y)^2  \right]Y(\phi^I,x^A) = \la\, Y(\phi^I,x^A)
\end{equation}
for some separation constant $\la$.  We can use the left hand side to define an operator $\Ocal{0}$ acting on scalar fields on $\Hcal$, whose properties are discussed in Section \ref{sec:decomposition}.

In the gravitational case $b=s=2$, and inserting the terms given above into (\ref{eqn:gravperts}) allows us to define an operator $\Ocal{2}$ by
\begin{equation}
  Y_{\mu\nu}\left[(D^2 - q^2)\chi_2\right] =  \\ \chi_2 (\Ocal{2} Y)_{\mu\nu}
\end{equation}
The operator $\Ocal{2}$ obtained in this way is given explicitly by (\ref{eqn:Ograv}).  Proving that this operator is self-adjoint with respect to the inner product (\ref{eqn:gravinner}) given is a now a case of integrating by parts.

Similarly, for electromagnetic perturbations, $b=s=1$, and inserting the terms given into (\ref{eqn:maxperts}) give us the operator (\ref{eqn:Omax}).

\section{Computations for MP black holes with equal angular momenta}\label{app:technical}

In this section, we explain in detail how to obtain the results described in Section \ref{sec:nhmp}.

\subsection{Near-horizon geometry of extremal MP black holes}\label{app:nhmp}
Consider the near-horizon geometry of an extremal Myers-Perry black hole, described by the metric \eqref{eqn:metric}.  Given the results of Section \ref{sec:decoupling}, it suffices to study the $(d-2)$-dimensional space $\Hcal$.  We work in a frame
\begin{equation}\label{eqn:Hframe}
  e_2 = B(d\psi + \Acal), \quad
  e_\alh = r_+ \hat{e}_\alh ,
\end{equation}
where $\hat{e}_\alh$ are a real, orthonormal frame for $\CP{N}$, and $\Acal = \Acal_\alh e_\alh$.  With indices raised this gives
\begin{equation}
  e_2 = \frac{1}{B}\frac{\pd}{\pd \psi}, \quad
  e_\alh = \frac{1}{r_+} \left(\hat{e}_\alh - \Acal_\alh \frac{\pd}{\pd \psi}\right),
\end{equation}
Note that these vectors satisfy $e_i.e_j = \del_{ij}$, where $i,j,\ldots$ are frame basis indices on $\Hcal$.

The spin connection 1-forms $\om_{ij}$ associated with this basis are (recalling $E = 2L^2 / (B\Om)$):
\begin{equation}
  \om_{2\alh}     = \frac{B}{r_+^2} \Jcalh_{\alh\betah} e_\betah, \qquad
  \om_{\alh\betah} = -\frac{B}{r_+^2} \Jcalh_{\alh\betah} e_2 
                   + \frac{1}{r_+} \hat{\om}_{\alh\betah}.\label{eqn:curv1forms}
\end{equation}
where $\hat{\om}$ is the spin connection for $\CP{N}$, and $\Jcal = \frac{1}{2} \Jcalh_{\alh\betah}\hat{e}_\alh\hat{e}_\betah$ are the components of the complex structure for $\CP{N}$.  The resulting curvature 2-forms are
\begin{equation}
  \Rcal_{2\alh}   = \frac{B^2}{r_+^4} \del_{\alh\betah} e_2 \wedge e_\betah \eqand
  \Rcal_{\alh\betah} = \frac{1}{r_+^2} \hat{\Rcal}_{\alh\betah}
                      - \frac{B^2}{r_+^4} (\Jcalh_{\alh\betah}\Jcalh_{\gammah\delh} 
                      + \Jcalh_{\alh[\gammah|}\Jcalh_{\betah|\delh]}) e_\gammah\wedge e_\delh
\end{equation}
where $\hat{\Rcal}_{\alh\betah}$ are the curvature 2-forms on $\CP{N}$.

This results in a Riemann tensor with non-vanishing components
\begin{equation}\label{eqn:riemann}
  R_{2\alh 2\betah} = \frac{B^2}{r_+^4} \del_{\alh\betah} , \qquad
  R_{\alh\betah\gammah\delh} = \frac{1}{r_+^2} \hat{R}_{\alh\betah\gammah\delh} 
                              - \frac{2B^2}{r_+^4} (\Jcalh_{\alh\betah}\Jcalh_{\gammah\delh} 
                              + \Jcalh_{\alh[\gammah|}\Jcalh_{\betah|\delh]}).
\end{equation}
where 
\begin{equation}
  \Rh_{\al\beta\gamma\del} = \gh_{\al\gamma}\gh_{\beta\del} - \gh_{\al\del}\gh_{\beta\gamma}
                                 + \Jcalh_{\al\gamma}\Jcalh_{\beta\del} - \Jcalh_{\al\del}\Jcalh_{\beta\gamma} + 2\Jcalh_{\al\beta} \Jcalh_{\gamma\del}
\end{equation}
is the Riemann tensor of $\CP{N}$.  The non-vanishing Ricci tensor components and Ricci scalar are
\begin{equation}\label{eqn:ricci}
  R_{22} = \frac{2NB^2}{r_+^4}, \qquad
  R_{\alh\betah} = \left( \frac{2(N+1)}{r_+^2} - \frac{2B^2}{r_+^4} \right)\del_{\alh\betah}, \qquad
  R = \frac{4N(N+1)}{r_+^2} - \frac{2NB^2}{r_+^4}
\end{equation}

Note that the Einstein equations for the metric (\ref{eqn:metric}) are equivalent to the following algebraic relations:
\begin{equation}\label{id:einstein}
  \La = \frac{2}{E^2} - \frac{1}{L^2} 
      = -\frac{2}{E^2} + \frac{2NB^2}{r_+^4}
      =  \frac{2(N+1)}{r_+^2} - \frac{2B^2}{r_+^4} 
\end{equation}
These are solved automatically by equations (\ref{eqn:Ldef}-\ref{eqn:Edef}), but these relations are often useful for simplifying calculations.
%

When $\La=0$ (or equivalently $l\rightarrow \infty$), the full spacetime is asymptotically flat, and the identities (\ref{id:einstein}) simplify to
\begin{equation} \label{eqn:asflateinstein}
  E^2 = 2L^2 = \frac{B^2}{N(N+1)^2} = \frac{r_+^2}{N(N+1)}.
\end{equation}

\subsection{Computing perturbation operators for this example}
In Section \ref{sec:decoupling}, and the associated Appendix \ref{app:nhframe}, we derived equations that are covariant on $\Hcal$, with indices $\mu,\nu,\dots$.  This is convenient, in that it now allows us to evaluate these equations without using the particular basis choice that we used to derive them.

Here, $\Hcal$ can be written as a fibration over $\CP{N}$.  It will be convenient in this section to write equations in a way that is covariant over $\CP{N}$; since this will then allow us to divide components up into scalar, vector and tensor parts, depending on how they transform as fields on $\CP{N}$.  We define indices $\al,\beta,\ldots$ that are covariant on $\CP{N}$, raised and lowered with the Fubini-Study metric $\gh_{\al\beta}$ on $\CP{N}$.

For quantities transforming as vectors on $\CP{N}$, it is often useful to project into the $\mp i$ eigenspaces of $\Jcalh$ using the operator
\begin{equation}
  \Pcalh^\pm_{\al\beta} = \frac{1}{2} \left( \gh_{\al\beta} \pm i \Jcalh_{\al\beta}\right).
\end{equation}

We now look to evaluate the perturbation operators $\Ocal{b}$ in the case of this metric, using equations (\ref{eqn:Oscalar},\ref{eqn:Ograv},\ref{eqn:Omax}).  Here, $L$ is constant, so $d(L^2) = 0$ and various terms vanish.  Furthermore, (\ref{eqn:nhmpk}) implies that the vector field $k$ satisfies
\begin{equation}\label{eqn:kforms}
  k = \Om B e_2, \qquad
  k.m = \Om m, \qquad
  k.k = B^2\Om^2 ,\qquad 
  dk = \Om B \, de_2 = 2 \Om B^2 \Jcalh .
\end{equation}
Finally, we need to expand the covariant derivative on $\Hcal$ in terms of derivatives on $\CP{N}$.  It is convenient to define the following charged covariant derivative on $\CP{N}$:
\begin{equation}
   \Dcalh_\al = \hat{D}_\al - im \Acalh_\al,
\end{equation}
where $\Jcalh = \tfrac{1}{2}d\Acalh$ is the K\"ahler form on $\CP{N}$, and $\hat{D}_\al$ is the Levi-Civita connection.  
Note that, acting on a scalar, the charged derivative $\Dcalh$ satisfies
\begin{equation}
  [\Dcalh_\al,\Dcalh_\beta] = -2im \Jcalh_{\al\beta} \eqand
  \Dcalh^\pm . \Dcalh^\mp = \tfrac{1}{2} \Dcalh^2 \mp 2mN,
\end{equation}
where $\Dcalh^\pm_\al \equiv \Pcalh^{\pm\beta}_{\al}\Dcalh_\beta$.

Given this, we can expand terms of the form $\nablah^2 Y$ and $\nablah Y$ in terms of this derivative, some examples of components in the gravitational case include
\begin{equation}
  (\nablah^2 Y)_{22} = \left(\frac{1}{r_+^2} \Dcalh^2 -\frac{m^2}{B^2} - \frac{2(2N+1)B^2}{r_+^4} \right) Y_{22}
                      -\frac{4B}{r_+^3} \Jcalh^{\al\beta} \Dcalh_\al Y_{2\beta}
\end{equation}
and
\begin{equation}
  \nablah^\al Y_{2\al} = \frac{1}{r_+} \Dcalh^\al Y_{2\al}.
\end{equation}

Putting these expressions, together with equations (\ref{eqn:riemann},\ref{eqn:ricci},\ref{eqn:kforms}) into the general equations (\ref{eqn:Oscalar},\ref{eqn:Omax},\ref{eqn:Ograv}) gives us explicit expressions for the operators $\Ocal{0}$, $\Ocal{1}$ and $\Ocal{2}$ in the case of this metric.  The explicit expressions for these operators can then be simplified to those given in (\ref{eqn:2maxpert}-\ref{eqn:amaxpert}) for $\Ocal{1}$ and (\ref{eqn:22pert}-\ref{eqn:abpert}) for $\Ocal{2}$.

\subsection{Mode decomposition of operators}
We now move on to consider the more complicated case of Maxwell and gravitational perturbations.  Firstly, it is useful decompose the action of the operators $\Ocal{1}$ and $\Ocal{2}$ on an arbitrary eigenvector $Y$ into components tangent, and normal to, $\CP{N}$.

The operator $\Ocal{1}$ describing Maxwell perturbation modes (defined in equation (\ref{eqn:Omax})) reduces to
\begin{equation}\label{eqn:2maxpert}
  (\Ocal{1} Y)_{2} = \left( -\frac{2Nm^2L^4}{r_+^4} - \frac{L^2}{r_+^2} \Dcalh^2 + 2 + 4\La L^2
                     \right) Y_2 + 2\xi^{\al\beta}\Dcalh_\beta Y_\al
\end{equation}
and
\begin{equation}\label{eqn:amaxpert}
  (\Ocal{1} Y)_\al = \left( -\frac{2Nm^2L^4}{r_+^4} - \frac{L^2}{r_+^2} \Dcalh^2 
                           + \frac{2B^2L^2}{r_+^4} + \La L^2\right) Y_\al
        + \frac{2imL^2}{r_+^2} \left( \Jcalh_{\al}^{\;\;\beta} Y_{\beta} \right)
        - \xi_{\al}^{\phantom{\al}\beta}\Dcalh_\beta Y_2.
\end{equation}
where
\begin{equation}
  \xi_{\al\beta} \equiv \frac{L^2}{r_+} \left(\frac{1}{E} \gh_{\al\beta}
                                   - \frac{B}{r_+^2}\Jcalh_{\al\beta} \right).
\end{equation}
Indices in these equations are raised and lowered with the metric $\gh_{\al\beta}$ on $\CP{N}$.

It is also useful to define $\DelLA$; a charged Lichnerowicz operator acting on rank-2 symmetric tensors on $\CP{N}$:
\begin{equation}\label{eqn:Lichnerowicz}
  \DelLA \Ybb_{\al\beta} 
       = -\Dcalh^2 \Ybb_{\al\beta} - 2 \hat{R}_{\al\gamma\beta\del} \Ybb^{\gamma\del} 
         + 4(N+1) \Ybb_{\al\beta}.
\end{equation}
This is the obvious generalization of the standard Lichnerowicz operator on $\CP{N}$, with the Laplacian $\hat{\nabla}^2$ replaced by our charged Laplacian $\Dcalh^2$ (following \cite{Kunduri:2006}).

Given this definition, the action of the operator $\Ocal{2}$ for gravitational perturbations (\ref{eqn:Ograv}) on an arbitrary 2-tensor with Fourier dependence $e^{im\psi}$ is given by:
\begin{multline}\label{eqn:22pert}
  (\Ocal{2} Y)_{22} 
     = \left(-\frac{2Nm^2L^4}{r_+^4} + 2
        - \frac{L^2}{r_+^2} \Dcalh^2 + 4(N+1)\frac{L^2B^2}{r_+^4} -\frac{4(N+1)L^2}{l^2} 
       \right) Y_{22} \\
        + 4\xi^{\al\beta} \Dcal_\beta Y_{2\al},
\end{multline}
\begin{multline}\label{eqn:2bpert}
  (\Ocal{2} Y)_{2\al} 
       = \left(-\frac{2Nm^2L^4}{r_+^4} + 2
               - \frac{L^2}{r_+^2} \Dcalh^2
               - \frac{2L^2}{E^2} + (2N+6)\frac{B^2 L^2}{r_+^4} -\frac{4(N+1)L^2}{l^2} 
         \right) Y_{2\al} \\
        + \frac{2imL^2}{r_+^2} \Jcalh_\al^{\;\;\beta}Y_{2\beta}
        - 2\xi_{\;\;\al}^{\beta} \Dcalh_\beta Y_{22}
        + 2\xi^{\beta\gamma}\Dcalh_\gamma Y_{\al\beta},
\end{multline}
\begin{multline}\label{eqn:abpert}
  (\Ocal{2} Y)_{\al\beta} 
    = \left( -\frac{2Nm^2L^4}{r_+^4} - \frac{4(N+1)L^2}{r_+^2} 
             + \frac{4B^2 L^2}{r_+^4} \right)Y_{\al\beta} 
      + \frac{L^2}{r_+^2} \DelLA Y_{\al\beta} + \frac{2imL^2}{r_+^2}[\Jcalh,Y]_{\al\beta}\\
    - \frac{4B^2L^2}{r_+^4}\left( (\Jcalh Y \Jcalh)_{\al\beta} + \del_{\al\beta} Y_{22} \right)
    -4\xi_{(\al|}^{\quad \gamma} \Dcal_{\gamma} Y_{2|\beta)} ,
\end{multline}
Note that (\ref{eqn:22pert}) is equivalent to the trace of (\ref{eqn:abpert}), given that $Y_{22} = -Y_{\al}^{\;\;\al}$.

Recall that in Ref.\ \cite{decoupling}, we found decoupled equations for the quantities $\vphi_i$ and $\Om_{ij}$, and then in Section \ref{sec:decomposition} we separated each equation into an $AdS_2$ part and a part on $\Hcal\sim S^{2N+1}$.  In this example, we now see that there is further coupling that we want to get rid of, between equations on the different parts of $\Hcal$, namely the directions normal and tangent to $\CP{N}$.

We now look to complete the decoupling by taking a scalar-vector-tensor decomposition with respect to $\CP{N}$.  Our decomposition is equivalent to that used in the numerical studies of perturbations of the full spacetime \cite{Kunduri:2006,Murata:2008,Dias:2010eu}.  The result of this is that we can expand general perturbations in terms of scalar, vector and tensor harmonics on $\CP{N}$, and the relevant eigenvalues of the Laplacian $\Dcalh^2$ are known (see \cite{Martin:2008pf} for further details).  We describe this in detail below.  

Note that, for $N=1$, there are no vector or tensor modes.  That is, imposing either the conditions \eqref{eqn:TTdef} or the conditions \eqref{eqn:vecdef} implies that $Y_{\mu\nu}=0$.

\subsection{Gravitational Tensor modes}\label{sec:TTmodes}
Tensor modes are those that only have transverse, traceless parts of $\Om_{\al\beta}$ turned on, i.e.\ perturbations of the form
\begin{equation}\label{eqn:TTdef}
   Y_{22} = 0 = Y_{2\al}, \qquad 
  \gh^{\al\beta} Y_{\al\beta}=0, \qquad
  \Dcalh^{\pm\al} Y_{\al\beta} = 0. 
\end{equation}
The components of the equations (\ref{eqn:22pert},\ref{eqn:2bpert}) vanish for tensor type perturbations, and \eqref{eqn:abpert} reduces to
\begin{multline}\label{eqn:abpertTT}
  (\Ocal{2} Y)_{\al\beta}
    = \left( -\frac{2Nm^2L^4}{r_+^4} - \frac{4(N+1)L^2}{r_+^2} 
            + \frac{4B^2L^2}{r_+^4} \right)Y_{\al\beta} + \frac{L^2}{r_+^2} \DelLA Y_{\al\beta}
     + \frac{2imL^2}{r_+^2}[\Jcalh,Y]_{\al\beta}\\
    - \frac{4B^2L^2}{r_+^4}(\Jcalh Y \Jcalh)_{\al\beta},
\end{multline}
We expand $Y_{\al\beta}$ in terms of separable Fourier modes
\begin{equation}
  Y_{\al\beta} = e^{im\psi} \Ybb_{\al\beta}
\end{equation}
where $\Ybb_{\al\beta}(x)$ a tensor harmonic on $\CP{N}$,with $\Dcalh^{\al\pm} \Ybb_{\al\beta} = 0$.

As $\CP{N}$ is a complex manifold, we can split both $\Ybb$ and equation (\ref{eqn:abpertTT}) into hermitian and anti-hermitian parts, which are eigenvectors of the linear map
\begin{equation}\label{map:Jsquared}
  \Ybb_{\alpha\beta} \mapsto \Jcalh_{\al}^{\;\gamma} \Jcalh_{\beta}^{\;\del} \Ybb_{\gamma\del} 
\end{equation}
with eigenvalues $+1$ and $-1$ respectively.  In other words, we write $\Ybb_{\al\beta} = \Ybb^+_{\al\beta} + \Ybb^-_{\al\beta}$ where $(\Jcalh\Ybb^\pm \Jcalh)_{\al\beta} = \mp \Ybb^\pm_{\al\beta}$, with the upper signs corresponding to hermitian modes.

In the anti-hermitian case, the modes can be divided further into the $\mp i$ eigenspaces of $\Jcalh$, with $\Jcalh_{\al\beta} \Ybb^{\pm}_\beta = \mp i \Ybb^{\pm}_\beta$.  Following \cite{Kunduri:2006}, we summarize this by setting $\sigma = \mp 1$ ($-$ for hermitian, $+$ for anti-hermitian), and $\eps = \pm 1$ for the two cases of anti-hermitian modes, and then see that
\begin{equation}
  (\Jcalh\Ybb \Jcalh)_{\al\beta} = \sigma \Ybb_{\al\beta} \eqand [\Jcalh,\Ybb]_{\al\beta} 
                           = i\eps(1+\sigma) \Ybb_{\al\beta}.
\end{equation}

We can take $\Ybb$ to be an eigenstate of the generalized Lichnerowicz operator on $\CP{N}$ (as such eigenstates form a complete set), i.e. we assume that
\begin{equation}\label{eqn:evals}
  (\DelLA \Ybb)_{\al\beta} = \la_{\kap,m}^{\mathrm{T}} \Ybb_{\al\beta}.
\end{equation}

This eigenvalue equation has known solutions, discussed in \cite{Kunduri:2006}.  For $N=1$, there are no tensor harmonics on $\CP{1}=S^2$. For $N\geq 2$, the $m=0$ eigenvalues are given by
\begin{equation} \label{eqn:tensorevals}
  \lakz{T}= 4\kap(\kap + N) + 4(N+\sig),
\end{equation}
for non-negative integers $\kap$.\footnote{Note that the allowed range of values for $\kap$ is unknown in general, e.g.\ there may be a positive lower bound on the allowed values of $\kap$ in some dimensions, but this will not turn out to be relevant here.} 

Inserting all this into (\ref{eqn:abpertTT}) implies that
\begin{equation}
  (\Ocal{2} Y)_{\al\beta} = \la Y_{\al\beta}
\end{equation}
where 
\begin{equation}\label{eqn:TTreqn}
  \la  =  -\frac{2Nm^2L^4}{r_+^4}
            + \frac{4B^2L^2}{r_+^4}(1-\sig) 
            + (\la_{\kap,m}^{\mathrm{T}} - 4(N+1) - 2m(1+\sig))\frac{L^2}{r_+^2},
\end{equation}
In Section \ref{sec:nhmp} we gave this eigenvalue explicity in the asymptotically flat case \eqref{eqn:tensorevalsflat} and the asymptotically $AdS$ case \eqref{eqn:tensorevalsads}. 

\subsection{Gravitational Vector Modes}\label{sec:gravvec}
There have currently been no studies in the literature of the stability of this black hole to vector type perturbations, which exist in dimensions $d\geq 7$.

Vector modes consist of divergence-free vectors $Y_{2\al}$, along with the traceless, but not transverse, contributions to $Y_{\al\beta}$ that can be constructed from them by differentiation, that is
\begin{equation}\label{eqn:vecdef}
  Y_{22} = 0, \qquad 
  \Dcalh^{\pm\al} Y_{2\al} = 0,\qquad
  Y_{\al}^{\;\;\al} = 0.
\end{equation}
We expand these perturbations as
\begin{equation}
  Y_{2\al} = g e^{im\psi} \Ybb_\al, \qquad
  Y_{\al\beta} = e^{im\psi} \left(h^+ \Ybb_{\al\beta}^+ + h^- \Ybb_{\al\beta}^-\right)
               \equiv Y_{\al\beta}^+ + Y_{\al\beta}^-
\end{equation}
where $\Ybb_\al$ is a divergence-free vector harmonic with
\begin{equation} \label{eqn:vecevals}
  \Dcalh^2 \Ybb_\al = - \lak{V} \Ybb_\al, \qquad
  \Dcalh^{\pm\al} \Ybb_{\al} = 0, \eqand
  \Ybb_{\al\beta}^\pm \equiv \frac{-1 }{\sqrt{\lak{V}}} \Dcalh^\pm_{(\al}\Ybb_{\beta)}. 
\end{equation}

There are several different separable modes that couple to each other in this sector of perturbations.  Therefore, in order to find the relevant eigenvalues we need to consider all such modes together.  In particular, the eigenvalues of $\Ocal{2}$ will be the eigenvalues of the matrix that describes the coupling between the different components of $Y_{ij}$.

We can take $\Ybb_\al$ to be an eigenvector of the complex structure $\Jcalh$, with eigenvalue $i\eps = \mp i$, that is: $\Jcalh_{\al}^{\;\;\beta} \Ybb_\beta = -i\eps \Ybb_\al$.

Note that $\Ybb_{\al\beta}=0$ is traceless,
\begin{eqnarray}
 \Dcalh^2 Y_{\al\beta} 
    &=& - \left[\lak{V}-2(N+1) - 4m - 2(1+3\eps) 
                                       \right] Y_{\al\beta}^+ \nn\\
    & & - \left[\lak{V} -2(N+1)+ 4m - 2(1-3\eps) 
                                          \right] Y_{\al\beta}^-
\end{eqnarray}
and
\begin{equation}
  \Dcalh^{\pm\beta} Y_{\al\beta} = \frac{e^{im\psi}}{2\sqrt{\lak{V}}}
        \left[ \frac{\lak{V}}{2} \pm m(N+1\mp \eps) - (1\mp 2\eps)(N+1) \right]h^\mp \Ybb_\al.
\end{equation}

The action of $\Ocal{2}$ on $Y$ now reduces to three equations:
\begin{eqnarray}\label{eqn:2bpertvec}
  (\Ocal{2}Y)_{2\al}
      &=& \Bigg[ L^2\left(
                -\frac{2Nm^2L^2}{r_+^4} + \frac{2m\eps}{r_+^2} 
                + \frac{\lak{V}}{r_+^2}
                + \frac{2}{E^2} + (2N+6)\frac{B^2}{r_+^4}
          \right) g \nn\\
      & & \qquad + \frac{\xi^+}{\sqrt{\lak{V}}} \left( \tfrac{1}{2}\lak{V} + m(N+1-\eps)
                                               - (1-2\eps)(N+1) \right) h^- \\
      & & \qquad + \frac{\xi^-}{\sqrt{\lak{V}}} \left(\tfrac{1}{2}\lak{V} - m(N+1+\eps) 
                 - (1+2\eps)(N+1) \right) h^+  \Bigg] e^{im\psi}\Ybb_{\al}\nn
\end{eqnarray}
and
\begin{multline}\label{eqn:abpertvec}
  (\Ocal{2}Y)_{\al\beta}^\pm
    = \Bigg[ 
      L^2\bigg( -\frac{2Nm^2L^2}{r_+^4} \mp \frac{2m}{r_+^2} + \frac{2m\eps}{r_+^2}
             + \frac{4B^2}{r_+^4}(1\mp \eps) - \frac{2(N+1)}{r_+^2} 
             + \frac{\lak{V}}{r_+^2}
      \bigg) h^\pm \\
     + 4\sqrt{\lak{V}} \xi^\pm g \Bigg]e^{im\psi}\Ybb_{\al\beta}^\pm,
\end{multline}
where
\begin{equation}
  \xi^\pm \equiv  \frac{L^2}{r_+} \left(\frac{1}{E} \pm \frac{iB}{r_+^2} \right), \qquad \xi \equiv \xi^+.
\end{equation}
To obtain the latter equation, we have separated out the components proportional to $\Ybb^\pm$ by noting that they are both eigenfunctions of the map \eqref{map:Jsquared} with differing eigenvalues $\pm\eps$.

Hence we have obtained a matrix formulation of the operator $\Ocal{2}$ in this case, acting on $[g,h^+,h^-]^\mathrm{T}$.  We can think of this as describing the mixing between the sectors $Y_\al$, $Y^+_{\al\beta}$, $Y^-_{\al\beta}$:
\begin{multline}
  \Ocal{2}
    = L^2 \left( \tfrac{\lak{V}+2m\eps}{r_+^2} -\tfrac{2Nm^2L^2}{r_+^4} \right) \Id \\
    +  \left( \begin{array}{ccc}
           \frac{2L^2}{E^2} +  \frac{(2N+6)L^2 B^2}{r_+^4} 
             & {\scriptstyle \tfrac{\left(\frac{1}{2}\lak{V} - m(N+1+\eps) 
                                               - (1+2\eps)(N+1) \right)\xi^*}{\sqrt{\lak{V}}} }
             & {\scriptstyle \tfrac{\left( \frac{1}{2}\lak{V} + m(N+1-\eps)
                                               - (1-2\eps)(N+1) \right)\xi}{\sqrt{\lak{V}}} }\\
           4\xi\sqrt{\lak{V}} & \frac{4B^2L^2}{r_+^4}(1-\eps) - \frac{2(N+1+m)L^2}{r_+^2} & 0 \\
           4\xi^*\sqrt{\lak{V}} & 0 & \frac{4B^2L^2}{r_+^4}(1+\eps) - \frac{2(N+1-m)L^2}{r_+^2}
        \end{array}\right)
\end{multline}

We now restrict to the case $m=0$ that is relevant to our conjecture, and find that here the matrix now reduces to
\begin{equation} \label{eqn:vecmatrix}
  \left( \begin{array}{ccc}
           \frac{\lakz{V}L^2}{r_+^2} + \frac{2L^2}{E^2} +\frac{(2N+6)B^2 L^2}{r_+^4} 
             & {\scriptstyle \left(\tfrac{1}{2}\lakz{V}
                                   - (1+2\eps)(N+1) \right)\tfrac{\xi^*}{\sqrt{\lakz{V}}}}
             & {\scriptstyle\left(\tfrac{1}{2}\lakz{V} 
                                  - (1-2\eps)(N+1)\right)\tfrac{\xi}{\sqrt{\lakz{V}}}}\\
           4\xi\sqrt{\lakz{V}} & \frac{L^2(\lakz{V}-2(N+1))}{r_+^2} + \frac{4B^2L^2}{r_+^4}(1-\eps)
             & 0 \\
           4\xi^*\sqrt{\lakz{V}} 
             & 0 
             & \frac{L^2(\lakz{V}-2(N+1))}{r_+^2} + \frac{4B^2L^2}{r_+^4}(1+\eps)
        \end{array}\right)
\end{equation}
We can find all eigenvalues of $\Ocal{2}$ by finding the eigenvalues of this matrix.  However, to do this explicitly we need to determine the allowed eigenvalues $\lakz{V}$ of $-\Dcalh^2 = -\hat{\nabla}^2$.  Note that the eigenvalues $\la_H$ of the Hodge-de Rham Laplacian
\begin{equation}
 \Del_H = -(\star d \star d + d \star d \star)
\end{equation}
on $\CP{3}$ were given in Ref.\ \cite[Table 2]{Martin:2008pf} (determined from \cite{Ikeda:1978}).  These can be generalized to $\CP{N}$ to give
\begin{equation}
  \la_H = 4(\kap+2)(\kap+N+1) \quad \mathrm{where} \quad \kap = 0,1,2,\ldots.
\end{equation}
The eigenvalues of the standard Laplacian are related to this by the Bochner-Weitzenb\"ock identity on $\CP{N}$, which implies that
\begin{equation}
  \Del_H \Ybb_\al = -\hat{\nabla}^2\Ybb_\al + 2(N+1) \Ybb_\al
\end{equation}
where we have made use of the Ricci tensor
\begin{equation}\label{eqn:cpnricci}
  \Rh_{\al\beta} = 2(N+1)\gh_{\al\beta}
\end{equation}
of $\CP{N}$.  Hence the eigenvalues of $-\hat{\nabla}^2$are actually
\begin{equation}\label{eqn:vecla}
  \lakz{V} =  4(\kap+2)(\kap+N+1) - 2(N+1) = 4\kap (\kap + 2) + 2(N+1)(2\kap+3)
\end{equation}
where $\kap = 0,1,2,\dots$.  This gives us enough information to evaluate the eigenvalues of $\Ocal{2}$.

In the asymptotically flat case the matrix representation of $\Ocal{2}$ reduces, using the identities \eqref{eqn:asflateinstein}, to
\begin{multline}\label{eqn:O2scalar}
  \frac{\Ocal{2}}{L^2} = \frac{\lakz{V}-2(N+1)}{r_+^2}\Id\\
      + \left( \begin{array}{ccc}
           \frac{2}{NL^2}(N+2) 
            &  \left(\tfrac{1}{2}\lakz{V}- (1+2\eps)(N+1) \right)\tfrac{\xi^*}{\sqrt{\lakz{V}}}
            & \left(\tfrac{1}{2}\lakz{V} - (1-2\eps)(N+1)\right)\tfrac{\xi}{\sqrt{\lakz{V}}}\\
           4\xi \sqrt{\lakz{V}}    & \frac{2}{NL^2}(1-\eps)  & 0 \\
           4\xi^* \sqrt{\lakz{V}}  &             0           
                                   & \frac{2}{NL^2}(1+\eps)
        \end{array}\right).
\end{multline}
The characteristic equation is then independent of $\eps$.  Inserting the allowed values \eqref{eqn:vecla} into this, we find that the eigenvalues of $\Ocal{2}$ are simple rational numbers, given by equation \eqref{eqn:gravvecevals}.

In the asymptotically $AdS$ case, it is not possible to find the eigenvalues explicitly (at least in a simple form).  However, it is reasonably straightforward to prove that all eigenvalues are positive for all $N$ and $\kap$, and hence there is no instability in this sector.

\subsection{Gravitational Scalar Modes}\label{sec:gravscalar}
Next, we consider the sector of scalar perturbations.  For the (non-extremal) full black hole solution, such perturbations have been previously studied by Murata \& Soda \cite{Murata:2008} (for $d=5$) and Dias \etal\ \cite{Dias:2010eu} (for $d=5,7,9$).

Scalar modes are the most complicated, with all possible parts of the perturbations turned on.  Starting with $Y_{22}$, contributions to $Y_{2\al}$ and $Y_{\al\beta}$ are constructed by taking derivatives.  Recall that the scalar eigenfunctions (\ref{eqn:scalarYbb}) of the charged covariant Laplacian $\Dcalh^2$ on $\CP{N}$ have eigenvalues given in (\ref{eqn:scalarevals}).  We can describe the full set of scalar perturbations as
\begin{align}
  Y_{22}          &= e^{im\psi} f \Ybb,\nn \\
  Y_{2\alpha}     &= e^{im\psi} \left[g^+ \Ybb_\al^+ + g^- \Ybb_\al^- \right],\nn\\
  Y_{\alpha\beta} 
      &= e^{im\psi}\bigg[- \tfrac{1}{\sqrt{\lak{S}}}\Big(h^{++} \Ybb_{\al\beta}^{++} 
                                                     + h^{--} \Ybb_{\al\beta}^{--} 
                                                     + h^{+-} \Ybb_{\al\beta}^{+-} \Big)
                           - \tfrac{1}{2N}f\del_{\al\beta} \Ybb \bigg],\label{eqn:scalaransatz}
\end{align}
where $\Ybb$ is the scalar eigenfunction defined in $(\ref{eqn:scalarYbb})$ and $\Ybb_\al^\pm$, $\Ybb_{\al\beta}^{\pm\pm}$, $\Ybb_{\al\beta}^{+-}$ are scalar-derived vector/tensor eigenfunctions, defined by
\begin{equation}\label{eqn:scalarderived}
  \Ybb_{\al}^\pm \equiv -\frac{\Dcalh_\al^\pm \Ybb}{\sqrt{\lak{S}}}  , \qquad
  \Ybb_{\al\beta}^{\pm\pm} \equiv \Dcalh^\pm_{(\al} \Ybb_{\beta)}^\pm 
\end{equation}
and
\begin{equation}
  \Ybb_{\al\beta}^{+-} = \Dcalh^+_{(\al} \Ybb_{\beta)}^- + \Dcalh^-_{(\al} \Ybb_{\beta)}^+ 
                           - \tfrac{\sqrt{\lak{S}}}{2N}\del_{\al\beta}\Ybb.
\end{equation}
These have the following properties:
\begin{align}
  \Jcalh_{\al}^{\;\;\beta} \Ybb^{\pm}_\beta &= \mp i \Ybb^\pm_\al, &
  \Dcalh^2 \Ybb_\al^\pm         &= -\left[\lak{S} - 2(N+1)\mp 4m \right] \Ybb_\al^\pm \nn\\
  \gh^{\al\beta} \Dcalh_\al \Ybb^\pm_{\beta} 
                                 &= \tfrac{ \lak{S} \mp 2mN}{2\sqrt{\lak{S}}} \Ybb ,&
   \Dcalh^2 \Ybb^{\pm\pm}_{\al\beta} &= -\left[\lak{S} - 4(N+3) \mp 8m \right] \Ybb^\pm_{\al\beta}, \nn\\
   \Jcalh^{\al\beta} \Dcalh_\al \Ybb^\pm_{\beta} 
                                 &= \tfrac{\mp i}{2\sqrt{\lak{S}}} \left( \lak{S} \mp 2mN \right) \Ybb ,&
   \Dcalh^2 \Ybb^{+-}_{\al\beta}     &= -\left(\lak{S} - 4N \right) \Ybb^{+-}_{\al\beta},
\end{align}
\begin{equation}
  (\Jcalh \Ybb^{\pm\pm}\Jcalh)_{\al\beta} = + \Ybb_{\al\beta}, \qquad
  (\Jcalh \Ybb^{+-}\Jcalh)_{\al\beta} = - \Ybb_{\al\beta},\qquad
  (\Jcalh\Ybb^{\pm\pm})_{\al\beta} = \mp i \Ybb_{\al\beta}
\end{equation}
and
\begin{align}
  \Dcalh^\beta \Ybb_{\al\beta}^{\pm\pm} 
                &= -\tfrac{1}{2}\left( \lak{S} - 4(N+1) \mp 2m(N+2) \right)\Ybb^\pm_\al,\nn\\
  \Dcalh^\beta \Ybb_{\al\beta}^{+-}
                &= -\tfrac{N-1}{2N}\left[ (\lak{S}+2mN)\Ybb^+_\al + (\lak{S}-2mN)\Ybb^-_\al \right],\nn\\
  \Jcalh^{\beta\gamma} \Dcalh_\gamma \Ybb_{\al\beta}^{\pm\pm} 
                &= \mp\tfrac{i}{2}\left[ \lak{S} - 4(N+1) \mp 2m(N+2) \right]\Ybb^\pm_\al,\nn\\
  \Jcalh^{\beta\gamma} \Dcalh_\gamma \Ybb_{\al\beta}^{+-} 
                &= \tfrac{i(N-1)}{2N}\left[ (\lak{S}+2mN)\Ybb^+_\al - (\lak{S}-2mN)\Ybb^-_\al \right].
\end{align} 
Note that there are three exceptions to this description:
\begin{itemize}
\item For $\kap = m = 0$, $\Ybb$ is constant, and there are no scalar-derived vectors or tensors.  Here the system is described by just one equation.
\item For $\kap=1, m=0$, the functions $\Ybb^{\pm\pm}$ vanish, and there are only four relevant types of component.
\item For $N=1$ (i.e.\ in five dimensions), the function $\Ybb^\pm$ vanishes identically (as there are no traceless, symmetric type (1,1) tensors on $\CP{1}$).
\end{itemize}

Inserting the ansatz \eqref{eqn:scalaransatz} into equations (\ref{eqn:22pert}-\ref{eqn:abpert}), we obtain the following.  From (\ref{eqn:22pert}) we get
\begin{multline} \label{eqn:f1}
  (\Ocal{2}Y)_{22} = \Bigg[\left( -\frac{2Nm^2L^4}{r_+^4} 
                     + \frac{4L^2}{E^2}  
                     + \frac{\lak{S}L^2}{r_+^2} + 4(N+1)\frac{B^2L^2}{r_+^4}\right) f\\
                     + \frac{2\xi^- (\lak{S} - 2mN)g^+}{\sqrt{\lak{S}}} 
                     + \frac{2\xi^+ (\lak{S} + 2mN)g^-}{\sqrt{\lak{S}}} \Bigg] e^{im\psi}\Ybb .
\end{multline}
Splitting (\ref{eqn:2bpert}) into $\mp i$ eigenspaces of $\Jcalh$ gives two equations
\begin{multline} \label{eqn:gpm1}
  (\Ocal{2}Y)_{2\al}^\pm
           = \Bigg[ \left( -\frac{2Nm^2L^4}{r_+^4} 
                       + \frac{\left(\lak{S} - 2(N+1) \mp 2m\right)L^2}{r_+^2}
                       + \frac{2L^2}{E^2} + (2N+6) \frac{B^2L^2}{r_+^4} \right) g^\pm \\
              + 2\sqrt{\lak{S}} \xi^\pm \left(1+\frac{1}{2N}\right) f 
              + \frac{\xi^\pm}{\sqrt{\lak{S}}} \left( \frac{N-1}{N}\right) (\lak{S} \pm 2mN) h^{+-} \\
              + \frac{\xi^\pm}{\sqrt{\lak{S}}} \left(\lak{S} - 4(N+1) \mp 2m(N+2)\right) h^{\pm\pm} 
             \Bigg] e^{im\psi}\Ybb_{\al}^\pm
\end{multline}
and from (\ref{eqn:abpert}) we obtain three equations
\begin{equation} \label{eqn:hpmpm1}
  (\Ocal{2}Y)_{\al\beta}^{\pm\pm}
      = \Bigg[ \left( -\frac{2Nm^2L^4}{r_+^4} 
                        + \frac{\left(\lak{S}-4(N+1)\mp 4m\right)L^2 }{r_+^2} \right) h^{\pm\pm} 
                   + 4\sqrt{\lak{S}}\xi^\pm g^\pm \Bigg] e^{im\psi}\Ybb_{\al\beta}^{\pm\pm}
\end{equation}
and

\begin{multline} \label{eqn:hpm1}
  (\Ocal{2}Y)_{\al\beta}^{+-} 
           =\Bigg[ \left( -\frac{2Nm^2L^4}{r_+^4} 
                            + \frac{\left(\lak{S} - 4(N+1)\right)L^2}{r_+^2} 
                           + \frac{8B^2L^2}{r_+^4} \right) h^{+-} \\
               + 2\sqrt{\lak{S}} (\xi^- g^+ + \xi^+ g^- ) \Bigg] e^{im\psi}\Ybb_{\al\beta}^{+-},
\end{multline}
as well as again obtaining (\ref{eqn:f1}) from the trace terms.

In a similar way to the vector case, we now get a matrix representation of $\Ocal{2}$, acting on $[f,g^+,g^-,h^{++},h^{--},h^{+-}]^\mathrm{T}$.  For simplicity, we display it explicitly here only in the case $m=0$:
\begin{equation}
 \tfrac{1}{L^2} \Ocal{2} = \tfrac{\lakz{S}-4(N+1)}{r_+^2} \Id + \nn
  \qquad\qquad\qquad\qquad \qquad\qquad\qquad\qquad\qquad\qquad\qquad\qquad\qquad\qquad\qquad
\end{equation}
{\tiny
\begin{equation}\label{eqn:scalarmat}
 \left(
      \begin{array}{cccccc}
        2\La + \frac{4(N+2)B^2}{r_+^4} + \tfrac{4}{E^2}
              & \tfrac{2\xi^* \sqrt{\lakz{S}}}{L^2} & \tfrac{2\xi  \sqrt{\lakz{S}}}{L^2} 
              & 0 & 0 & 0 \\[3mm]
       \frac{2\xi}{L^2} \sqrt{\lakz{S}}  \left(1+\frac{1}{2N}\right) 
              &  \La + \frac{2}{E^2} + \frac{2(N+4) B^2}{r_+^4}   & 0 
              &{\scriptstyle\frac{(\lakz{S} - 4(N+1))\xi^*}{L^2\sqrt{\lakz{S}}}} 
              & 0 & \frac{(N-1)\xi \sqrt{\lakz{S}}}{NL^2} \\[3mm]
      \frac{2\xi^*}{L^2}\sqrt{\lakz{S}} \left(1+\frac{1}{2N}\right) & 0 
              & \La + \frac{2}{E^2}  + \frac{2(N+4) B^2}{r_+^4} 
              & 0 & {\scriptstyle\frac{(\lakz{S} - 4(N+1))\xi}{L^2\sqrt{\lakz{S}}}}
              & \frac{(N-1)\xi^* \sqrt{\lakz{S}}}{NL^2} \\[3mm]
      0 & \frac{4\xi}{L^2} \sqrt{\lakz{S}} & 0 &  0 & 0 & 0 \\[3mm]
      0 & 0 & \frac{4\xi^*}{L^2} \sqrt{\lakz{S}} & 0 & 0 & 0 \\[3mm]      
      0 & \frac{2\xi^*}{L^2}\sqrt{\lakz{S}} & \frac{2\xi}{L^2}\sqrt{\lakz{S}} 
              & 0 & 0 & \frac{8B^2}{r_+^4}
      \end{array}
  \right)
\end{equation}}
Again, although this matrix is complex, its eigenvalues are all real, and we now look to compute these explicitly, using the list of scalar eigenvalues $\lak{S}$ of $\Dcalh^2$ given by \eqref{eqn:scalarevals}.

Recall from above the there are three special cases that need to be dealt with separately.

Firstly, the case $\kap=m=0 = \la_{0,0}^\mathrm{S}$ is degenerate, in the sense that $Y_{2\al}$ and $Y_{\al\beta}$ vanish.  Hence this matrix reduces to a $1\times 1$ matrix,
\begin{equation}
  (\Ocal{2} Y)_{22} = L^2\left( \frac{4}{E^2} + 4(N+1)\frac{B^2}{r_+^4}\right) Y_{22}
\end{equation}
which has a trivially positive eigenvalue.

When $m=0,\kap=1$,  $\la_{1,0}^\mathrm{S} =4(N+1)$ and the eigenfunctions $\Ybb^{\pm\pm}$ vanish, which means that the matrix representation of $\Ocal{2}$ is actually a $4\times4$ matrix, with
{\small
\begin{equation}
 \Ocal{2}=  L^2
 \left(
      \begin{array}{cccc}
        \frac{4(N+2)B^2}{r_+^4}+ 2\La + \frac{4}{E^2}
             & \frac{4\xi^*\sqrt{N+1}}{L^2}  
             & \frac{4\xi\sqrt{N+1}}{L^2} & 0 \\[3mm]
        \frac{4 \xi\sqrt{N+1}}{L^2} \left(1+\frac{1}{2N}\right) 
             &  \La + \frac{2}{E^2} + \frac{2(N+4) B^2}{r_+^4} & 0 
             & \frac{2(N-1)\xi\sqrt{N+1}}{NL^2} \\[3mm]
        \frac{4\xi^* \sqrt{N+1}}{L^2} \left(1+\frac{1}{2N}\right) 
             & 0 & \La + \frac{2}{E^2} + \frac{2(N+4) B^2}{r_+^4}  
             & \frac{2(N-1)\xi^*\sqrt{N+1}}{NL^2} \\[3mm]
       0 & \frac{4\xi^*\sqrt{N+1}}{L^2} & \frac{4\xi\sqrt{N+1}}{L^2} & \frac{8B^2}{r_+^4}
      \end{array}
 \right)
\end{equation} }
The eigenvalues of this matrix were analysed in Sections \ref{sec:gravscalarevalsflat} and \ref{sec:gravscalarevalsads} in the asymptotically flat and asymptotically $AdS$ cases respectively, along with the eigenvalues of the $6\times 6$ matrix \eqref{eqn:scalarmat} for the case $\kap\geq 2$.

Finally, consider the case $N=1$, for which $\Ybb^{+-}$ vanishes. This has the effect of eliminating the final row and column from the above matrices.

\subsection{Electromagnetic vector modes}
Following a similar approach to that of the gravitational case, we can obtain results for electromagnetic perturbations.

Note that we do not necessarily see all possible Maxwell perturbations with this approach, as perturbations that change $F_{ij}$ or $F$, but not $\vphi$ or $\vphi'$, cannot be analysed.  It is not clear whether there exist non-trivial perturbations with this property.\footnote{One can of course consider perturbations of $\vphi'$ rather than $\vphi$ by taking the prime of all equations above.  This has the effect of mapping $q \mapsto q^*$, $\chi\mapsto \chi^*$, $\eps\mapsto -\eps$ and $m\mapsto -m$, but leaves all results unchanged.}

The Maxwell perturbation modes can be divided into two categories which we will refer to as `vectors' and `scalars', according to their transformation properties on $\CP{N}$.  Vector modes are those that only have a divergence-free $\CP{N}$ part of $Y$ turned on, that is
\begin{equation}
  Y_2 = 0  \eqand \Dcalh^{\pm\al} Y_\al = 0.
\end{equation}

The simplest class of electromagnetic perturbations are the vector modes, which we can parametrize as
\begin{equation}
  Y_2 = 0, \qquad Y_\al = e^{im\psi} \Ybb_\al,
\end{equation}
where $\Ybb_\al$ are the divergence-free vector eigenfunctions of $\Dcalh^2$ defined by \eqref{eqn:vecevals} above.  The component $(\Ocal{1}Y)_2$ vanishes, and \eqref{eqn:amaxpert} reduces to
\begin{equation}
 (\Ocal{1}Y)_\al = \left[-\frac{2Nm^2L^4}{r_+^4} 
                         + \frac{\left(\lak{V}+2(N+1)+2m\eps\right)L^2}{r_+^2}
                   \right]Y_\al.
\end{equation}
This gives the eigenvalues described in Section \ref{sec:emvectors}.

\subsection{Electromagnetic scalar modes}
The $\CP{N}$ scalar modes are more complicated, as for vector and scalar eigenvalues in the gravitational case.  We can expand the perturbations as
\begin{equation}
  Y_2   = e^{im\psi} f \Ybb, \qquad 
  Y_\al = e^{im\psi}\left(g^+ \Ybb^+_\al + g^- \Ybb^-_\al \right)
\end{equation}
where $\Ybb$ are the scalar eigenfunctions defined in \eqref{eqn:scalarYbb}, and $\Ybb^\pm_\al$ the scalar-derived vectors defined in \eqref{eqn:scalarderived}.

Note that for $\kap = m = 0$, when $\lak{S} = 0$, the associated eigenfunction $\Ybb(x)$ is constant, and hence $Y_\al=0$.  In this case, the operator $\Ocal{1}$ has simple eigenvalues, given by equation \eqref{eqn:emevals0}.

For $\lak{S}>0$, we follow an analagous separation procedure to that of the gravitational case, and find that the effective $AdS_2$ masses of various modes are given by eigenvalues of the matrix 
\begin{equation} \label{eqnmaxscalarmatrix}
 \Ocal{1} = \left( \begin{array}{ccc}
           \frac{\lak{S}L^2}{r_+^2} + 2 +4\La L^2
             &\tfrac{ (\lak{S}-2mN)\xi^*}{\sqrt{\lak{S}}}
             & \tfrac{(\lak{S}+2mN)\xi}{\sqrt{\lak{S}}}\\
           2\sqrt{\lak{S}}\xi & \frac{L^2(\lak{S}-2m)}{r_+^2}  & 0 \\
           2\sqrt{\lak{S}}\xi^* & 0 & \frac{L^2(\lak{S} +2m)}{r_+^2} 
        \end{array}\right).
\end{equation}
In the case $m=0$, the characteristic equation reduces to
\begin{equation}\label{eqn:maxwellcheqn}
  \left( \tfrac{L^2}{r_+^2}\lakz{S} - t\right)
  \Bigg[ t^2 - 2\left( \tfrac{L^2}{r_+^2}\lakz{S} + 1 + 2\La L^2 \right)t
         + \lakz{S} \left( \tfrac{L^4}{r_+^4} \lakz{S}  
                          + \tfrac{2L^2\left(1+2\La L^2 \right)}{r_+^2} - 4|\xi|^2\right) \Bigg] = 0,
\end{equation}
with allowed values of $\lakz{S}$ given by $\lakz{S} = 4\kap(\kap+N)$ for $\kap = 0,1,\ldots$.  This leads to the eigenvalues listed in Section \ref{sec:emscalars}.
%

\end{document}